# ARTIFICIAL INTELLIGENCE AND LIFE IN 2030

ONE HUNDRED YEAR STUDY ON ARTIFICIAL INTELLIGENCE | REPORT OF THE 2015 STUDY PANEL | SEPTEMBER 2016

## PREFACE

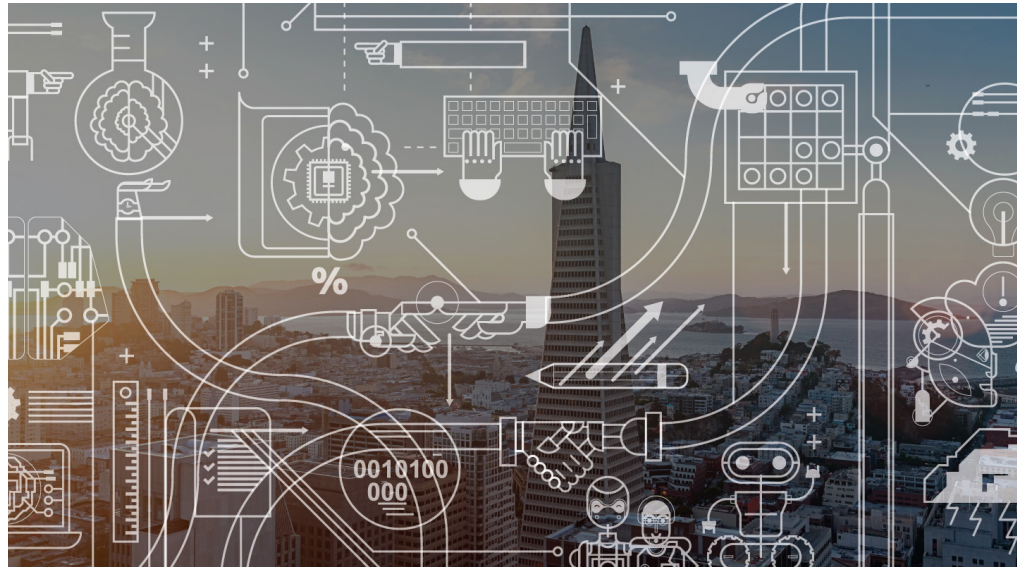

The One Hundred Year Study on Artificial Intelligence, launched in the fall of 2014, is a long-term investigation of the field of Artificial Intelligence (AI) and its influences on people, their communities, and society. It considers the science, engineering, and deployment of AI-enabled computing systems. As its core activity, the Standing Committee that oversees the One Hundred Year Study forms a Study Panel every five years to assess the current state of AI. The Study Panel reviews AI's progress in the years following the immediately prior report, envisions the potential advances that lie ahead, and describes the technical and societal challenges and opportunities these advances raise, including in such arenas as ethics, economics, and the design of systems compatible with human cognition. The overarching purpose of the One Hundred Year Study's periodic expert review is to provide a collected and connected set of reflections about AI and its influences as the field advances. The studies are expected to develop syntheses and assessments that provide expert-informed guidance for directions in AI research, development, and systems design, as well as programs and policies to help ensure that these systems broadly benefit individuals and society.[1]

The One Hundred Year Study is modeled on an earlier effort informally known as the "AAAI Asilomar Study." During 2008-2009, the then president of the Association for the Advancement of Artificial Intelligence (AAAI), Eric Horvitz, assembled a group of AI experts from multiple institutions and areas of the field, along with scholars of cognitive science, philosophy, and law. Working in distributed subgroups, the participants addressed near-term AI developments, long-term possibilities, and legal and ethical concerns, and then came together in a three-day meeting at Asilomar to share and discuss their findings. A short written report on the intensive meeting discussions, amplified by the participants' subsequent discussions with other colleagues, generated widespread interest and debate in the field and beyond.

The impact of the Asilomar meeting, and important advances in AI that included AI algorithms and technologies starting to enter daily life around the globe, spurred the idea of a long-term recurring study of AI and its influence on people and society. The One Hundred Year Study was subsequently endowed at a university to enable

> **The overarching purpose of the One Hundred Year Study's periodic expert review is to provide a collected and connected set of reflections about AI and its influences as the field advances.**

---

[1] "One Hundred Year Study on Artificial Intelligence (AI100)," Stanford University, accessed August 1, 2016, https://ai100.stanford.edu.

extended deep thought and cross-disciplinary scholarly investigations that could inspire innovation and provide intelligent advice to government agencies and industry.

This report is the first in the planned series of studies that will continue for at least a hundred years. The Standing Committee defined a Study Panel charge for the inaugural Study Panel in the summer of 2015 and recruited Professor Peter Stone, at the University of Texas at Austin, to chair the panel. The seventeen-member Study Panel, comprised of experts in AI from academia, corporate laboratories and industry, and AI-savvy scholars in law, political science, policy, and economics, was launched in mid-fall 2015. The participants represent diverse specialties and geographic regions, genders, and career stages.

The Standing Committee extensively discussed ways to frame the Study Panel charge to consider both recent advances in AI and potential societal impacts on jobs, the environment, transportation, public safety, healthcare, community engagement, and government. The committee considered various ways to focus the study, including surveying subfields and their status, examining a particular technology such as machine learning or natural language processing, and studying particular application areas such as healthcare or transportation. The committee ultimately chose a thematic focus on "AI and Life in 2030" to recognize that AI's various uses and impacts will not occur independently of one another, or of a multitude of other societal and technological developments. Acknowledging the central role cities have played throughout most of human experience, the focus was narrowed to the large urban areas where most people live. The Standing Committee further narrowed the focus to a typical North American city in recognition of the great variability of urban settings and cultures around the world, and limits on the first Study Panel's efforts. The Standing Committee expects that the projections, assessments, and proactive guidance stemming from the study will have broader global relevance and is making plans for future studies to expand the scope of the project internationally.

## TABLE OF CONTENTS

As one consequence of the decision to focus on life in North American cities, military applications were deemed to be outside the scope of this initial report. This is not to minimize the importance of careful monitoring and deliberation about the implications of AI advances for defense and warfare, including potentially destabilizing developments and deployments.

The report is designed to address four intended audiences. For the general public, it aims to provide an accessible, scientifically and technologically accurate portrayal of the current state of AI and its potential. For industry, the report describes relevant technologies and legal and ethical challenges, and may help guide resource allocation. The report is also directed to local, national, and international governments to help them better plan for AI in governance. Finally, the report can help AI researchers, as well as their institutions and funders, to set priorities and consider the ethical and legal issues raised by AI research and its applications.

Given the unique nature of the One Hundred Year Study on AI, we expect that future generations of Standing Committees and Study Panels, as well as research scientists, policy experts, leaders in the private and public sectors, and the general public, will reflect on this assessment as they make new assessments of AI's future. We hope that this first effort in the series stretching out before us will be useful for both its failures and successes in accurately predicting the trajectory and influences of AI.

The Standing Committee is grateful to the members of the Study Panel for investing their expertise, perspectives, and significant time to the creation of this inaugural report. We especially thank Professor Peter Stone for agreeing to serve as chair of the study and for his wise, skillful, and dedicated leadership of the panel, its discussions, and creation of the report.

**Standing Committee of the One Hundred Year Study of Artificial Intelligence**

Barbara J. Grosz, *Chair*   Russ Altman   Eric Horvitz
Alan Mackworth   Tom Mitchell   Deirdre Mulligan   Yoav Shoham

## STUDY PANEL

**Peter Stone**, University of Texas at Austin, Chair
**Rodney Brooks**, Rethink Robotics
**Erik Brynjolfsson**, Massachussets Institute of Technology
**Ryan Calo**, University of Washington
**Oren Etzioni**, Allen Institute for AI
**Greg Hager**, Johns Hopkins University
**Julia Hirschberg**, Columbia University
**Shivaram Kalyanakrishnan**, Indian Institute of Technology Bombay
**Ece Kamar**, Microsoft Research
**Sarit Kraus**, Bar Ilan University
**Kevin Leyton-Brown**, University of British Columbia
**David Parkes**, Harvard University
**William Press**, University of Texas at Austin
**AnnaLee (Anno) Saxenian**, University of California, Berkeley
**Julie Shah**, Massachussets Institute of Technology
**Milind Tambe**, University of Southern California
**Astro Teller**, X

**Acknowledgments:** The members of the Study Panel gratefully acknowledge the support of and valuable input from the Standing Committee, especially the chair, Barbara Grosz, who handled with supreme grace the unenviable role of mediating between two large, very passionate committees. We also thank Kerry Tremain for his tireless and insightful input on the written product during the extensive editing and polishing process, which unquestionably strengthened the report considerably.

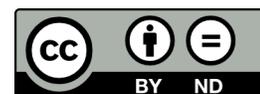

# EXECUTIVE SUMMARY

Artificial Intelligence (AI) is a science and a set of computational technologies that are inspired by—but typically operate quite differently from—the ways people use their nervous systems and bodies to sense, learn, reason, and take action. While the rate of progress in AI has been patchy and unpredictable, there have been significant advances since the field's inception sixty years ago. Once a mostly academic area of study, twenty-first century AI enables a constellation of mainstream technologies that are having a substantial impact on everyday lives. Computer vision and AI planning, for example, drive the video games that are now a bigger entertainment industry than Hollywood. Deep learning, a form of machine learning based on layered representations of variables referred to as neural networks, has made speech-understanding practical on our phones and in our kitchens, and its algorithms can be applied widely to an array of applications that rely on pattern recognition. Natural Language Processing (NLP) and knowledge representation and reasoning have enabled a machine to beat the Jeopardy champion and are bringing new power to Web searches.

> **Substantial increases in the future uses of AI applications, including more self-driving cars, healthcare diagnostics and targeted treatment, and physical assistance for elder care can be expected.**

While impressive, these technologies are highly tailored to particular tasks. Each application typically requires years of specialized research and careful, unique construction. In similarly targeted applications, substantial increases in the future uses of AI technologies, including more self-driving cars, healthcare diagnostics and targeted treatments, and physical assistance for elder care can be expected. AI and robotics will also be applied across the globe in industries struggling to attract younger workers, such as agriculture, food processing, fulfillment centers, and factories. They will facilitate delivery of online purchases through flying drones, self-driving trucks, or robots that can get up the stairs to the front door.

This report is the first in a series to be issued at regular intervals as a part of the One Hundred Year Study on Artificial Intelligence (AI100). Starting from a charge given by the AI100 Standing Committee to consider the likely influences of AI in a typical North American city by the year 2030, the 2015 Study Panel, comprising experts in AI and other relevant areas focused their attention on eight domains they considered most salient: transportation; service robots; healthcare; education; low-resource communities; public safety and security; employment and workplace; and entertainment. In each of these domains, the report both reflects on progress in the past fifteen years and anticipates developments in the coming fifteen years. Though drawing from a common source of research, each domain reflects different AI influences and challenges, such as the difficulty of creating safe and reliable hardware (transportation and service robots), the difficulty of smoothly interacting with human experts (healthcare and education), the challenge of gaining public trust (low-resource communities and public safety and security), the challenge of overcoming fears of marginalizing humans (employment and workplace), and the social and societal risk of diminishing interpersonal interactions (entertainment). The report begins with a reflection on what constitutes Artificial Intelligence, and concludes with recommendations concerning AI-related policy. These recommendations include accruing technical expertise about AI in government and devoting more resources—and removing impediments—to research on the fairness, security, privacy, and societal impacts of AI systems.

Contrary to the more fantastic predictions for AI in the popular press, the Study Panel found no cause for concern that AI is an imminent threat to humankind. No machines with self-sustaining long-term goals and intent have been developed, nor are they likely to be developed in the near future. Instead, increasingly useful applications of AI, with potentially profound positive impacts on our society and economy are likely to emerge between now and 2030, the period this report considers. At the same time, many of these developments will spur disruptions in how human labor is augmented or replaced by AI, creating new challenges for the economy and society more broadly. Application design and policy decisions made in the near term are likely to have long-lasting influences on the nature and directions of such developments, making it important for AI researchers, developers, social scientists, and policymakers to balance the imperative to innovate with mechanisms to ensure that AI's economic and social benefits are broadly shared across society. If society approaches these technologies primarily with fear and suspicion, missteps that slow AI's development or drive it underground will result, impeding important work on ensuring the safety and reliability of AI technologies. On the other hand, if society approaches AI with a more open mind, the technologies emerging from the field could profoundly transform society for the better in the coming decades.

**Study Panel: Peter Stone**, *Chair*, University of Texas at Austin, **Rodney Brooks**, Rethink Robotics, **Erik Brynjolfsson**, Massachussets Institute of Technology, **Ryan Calo**, University of Washington, **Oren Etzioni**, Allen Institute for AI, **Greg Hager**, Johns Hopkins University, **Julia Hirschberg**, Columbia University, **Shivaram Kalyanakrishnan**, Indian Institute of Technology Bombay, **Ece Kamar**, Microsoft Research, **Sarit Kraus**, Bar Ilan University. **Kevin Leyton-Brown**, University of British Columbia, **David Parkes**, Harvard University, **William Press**, University of Texas at Austin, **AnnaLee (Anno) Saxenian**, University of California, Berkeley, **Julie Shah**, Massachussets Institute of Technology, **Milind Tambe**, University of Southern California, **Astro Teller**, X

**Standing Committee of the One Hundred Year Study of Artificial Intelligence:**
Barbara J. Grosz, *Chair*, Russ Altman, Eric Horvitz, Alan Mackworth, Tom Mitchell, Deidre Mulligan, Yoav Shoham

> While drawing on common research and technologies, AI systems are specialized to accomplish particular tasks. Each application requires years of focused research and a careful, unique construction.

# OVERVIEW

The frightening, futurist portrayals of Artificial Intelligence that dominate films and novels, and shape the popular imagination, are fictional. In reality, AI is already changing our daily lives, almost entirely in ways that improve human health, safety, and productivity. Unlike in the movies, there is no race of superhuman robots on the horizon or probably even possible. And while the potential to abuse AI technologies must be acknowledged and addressed, their greater potential is, among other things, to make driving safer, help children learn, and extend and enhance people's lives. In fact, beneficial AI applications in schools, homes, and hospitals are already growing at an accelerated pace. Major research universities devote departments to AI studies, and technology companies such as Apple, Facebook, Google, IBM, and Microsoft spend heavily to explore AI applications they regard as critical to their futures. Even Hollywood uses AI technologies to bring its dystopian AI fantasies to the screen.

Innovations relying on computer-based vision, speech recognition, and Natural Language Processing have driven these changes, as have concurrent scientific and technological advances in related fields. AI is also changing how people interact with technology. Many people have already grown accustomed to touching and talking to their smart phones. People's future relationships with machines will become ever more nuanced, fluid, and personalized as AI systems learn to adapt to individual personalities and goals. These AI applications will help monitor people's well-being, alert them to risks ahead, and deliver services when needed or wanted. For example, in a mere fifteen years in a typical North American city—the time frame and scope of this report—AI applications are likely to transform transportation toward self-driving vehicles with on-time pickup and delivery of people and packages. This alone will reconfigure the urban landscape, as traffic jams and parking challenges become obsolete.

This study's focus on a typical North American city is deliberate and meant to highlight specific changes affecting the everyday lives of the millions of people who inhabit them. The Study Panel further narrowed its inquiry to eight domains where AI is already having or is projected to have the greatest impact: transportation, healthcare, education, low-resource communities, public safety and security, employment and workplace, home/service robots, and entertainment.

Though drawing from a common source of research, AI technologies have influenced and will continue to influence these domains differently. Each domain faces varied AI-related challenges, including the difficulty of creating safe and reliable hardware for sensing and effecting (transportation and service robots), the difficulty of smoothly interacting with human experts (healthcare and education), the challenge of gaining public trust (low-resource communities and public safety and security), the challenge of overcoming fears of marginalizing humans (employment and workplace) and the risk of diminishing interpersonal interaction (entertainment). Some domains are primarily business sectors, such as transportation and healthcare, while others are more oriented to consumers, such as entertainment and home service robots. Some cut across sectors, such as employment/workplace and low-resource communities.

In each domain, even as AI continues to deliver important benefits, it also raises important ethical and social issues, including privacy concerns. Robots and other AI technologies have already begun to displace jobs in some sectors. As a society, we are now at a crucial juncture in determining how to deploy AI-based technologies in ways that promote, not hinder, democratic values such as freedom, equality, and transparency. For individuals, the quality of the lives we lead and how our contributions are valued are likely to shift gradually, but markedly. Over the next several years, AI research, systems development, and social and regulatory frameworks will shape how the benefits of AI are weighed against its costs and risks, and how broadly these benefits are spread.

> **Many have already grown accustomed to touching and talking to their smart phones. People's future relationships with machines will become ever more nuanced, fluid, and personalized.**

An accurate and sophisticated picture of AI—one that competes with its popular portrayal—is hampered at the start by the difficulty of pinning down a precise definition of artificial intelligence. In the approaches the Study Panel considered, none suggest there is currently a "general purpose" AI. While drawing on common research and technologies, AI systems are specialized to accomplish particular tasks, and each application requires years of focused research and a careful, unique construction. As a result, progress is uneven within and among the eight domains.

A prime example is **Transportation**, where a few key technologies have catalyzed the widespread adoption of AI with astonishing speed. Autonomous transportation will soon be commonplace and, as most people's first experience with physically embodied AI systems, will strongly influence the public's perception of AI. As cars become better drivers than people, city-dwellers will own fewer cars, live further from work, and spend time differently, leading to an entirely new urban organization. In the typical North American city in 2030, physically embodied AI applications will not be limited to cars, but are likely to include trucks, flying vehicles, and personal robots. Improvements in safe and reliable hardware will spur innovation over the next fifteen years, as they will with **Home/Service Robots**, which have already entered people's houses, primarily in the form of vacuum cleaners. Better chips, low-cost 3D sensors, cloud-based machine learning, and advances in speech understanding will enhance future robots' services and their interactions with people. Special purpose robots will deliver packages, clean offices, and enhance security. But technical constraints and the high costs of reliable mechanical devices will continue to limit commercial opportunities to narrowly defined applications for the foreseeable future.

In **Healthcare**, there has been an immense forward leap in collecting useful data from personal monitoring devices and mobile apps, from electronic health records (EHR) in clinical settings and, to a lesser extent, from surgical robots designed to assist with medical procedures and service robots supporting hospital operations. AI-based applications could improve health outcomes and the quality of life for millions of people in the coming years. Though clinical applications have been slow to move from the computer science lab to the real-world, there are hopeful signs that the pace of innovation will improve. Advances in healthcare can be promoted via the development of incentives and mechanisms for sharing data and for removing overbearing policy, regulatory, and commercial obstacles. For many applications, AI systems will have to work closely with care providers and patients to gain their trust. Advances in how intelligent machines interact naturally with caregivers, patients, and patients' families are crucial.

Enabling more fluid interactions between people and promising AI technologies also remains a critical challenge in **Education**, which has seen considerable progress in the same period. Though quality education will always require active engagement by human teachers, AI promises to enhance education at all levels, especially by providing personalization at scale. Interactive machine tutors are now being matched to students for teaching science, math, language, and other disciplines. Natural Language Processing, machine learning, and crowdsourcing have boosted online learning and enabled teachers in higher education to multiply the size of their classrooms while addressing individual students' learning needs and styles. Over the next fifteen years in a typical North American city, the use of these technologies in the classroom and in the home is likely to expand significantly, provided they can be meaningfully integrated with face-to-face learning.

Beyond education, many opportunities exist for AI methods to assist **Low-resource Communities** by providing mitigations and solutions to a variety of social problems. Traditionally, funders have underinvested in AI research lacking commercial application. With targeted incentives and funding priorities,

> Society is now at a crucial juncture in determining how to deploy AI-based technologies in ways that promote rather than hinder democratic values such as freedom, equality, and transparency.

> Longer term, AI may be thought of as a radically different mechanism for wealth creation in which everyone should be entitled to a portion of the world's AI-produced treasures.

AI technologies could help address the needs of low-resource communities, and budding efforts are promising. Using data mining and machine learning, for example, AI has been used to create predictive models to help government agencies address issues such as prevention of lead poisoning in at-risk children and distribution of food efficiently. These budding efforts suggest more could be done, particularly if agencies and organizations can engage and build trust with these communities. Gaining public trust is also a challenge for AI use by **Public Safety and Security** professionals. North American cities and federal agencies have already begun to deploy AI technologies in border administration and law enforcement. By 2030, they will rely heavily upon them, including improved cameras and drones for surveillance, algorithms to detect financial fraud, and predictive policing. The latter raises the specter of innocent people being unjustifiably monitored, and care must be taken to avoid systematizing human bias and to protect civil liberties. Well-deployed AI prediction tools have the potential to provide new kinds of transparency about data and inferences, and may be applied to detect, remove, or reduce human bias, rather than reinforcing it.

Social and political decisions are likewise at play in AI's influences on **Employment and Workplace** trends, such as the safety nets needed to protect people from structural changes in the economy. AI is poised to replace people in certain kinds of jobs, such as in the driving of taxis and trucks. However, in many realms, AI will likely replace tasks rather than jobs in the near term, and will also create new kinds of jobs. But the new jobs that will emerge are harder to imagine in advance than the existing jobs that will likely be lost. AI will also lower the cost of many goods and services, effectively making everyone better off. Longer term, AI may be thought of as a radically different mechanism for wealth creation in which everyone should be entitled to a portion of the world's AI-produced treasures. It is not too soon for social debate on how the economic fruits of AI technologies should be shared.

**Entertainment** has been transformed by social networks and other platforms for sharing and browsing blogs, videos, and photos, which rely on techniques actively developed in NLP, information retrieval, image processing, crowdsourcing, and machine learning. Some traditional sources of entertainment have also embraced AI to compose music, create stage performances, and even to generate 3D scenes from natural language text. The enthusiasm with which people have already responded to AI-driven entertainment has been surprising. As with many aspects of AI, there is ongoing debate about the extent to which the technology replaces or enhances sociability. AI will increasingly enable entertainment that is more interactive, personalized, and engaging. Research should be directed toward understanding how to leverage these attributes for individuals' and society's benefit.

### What's next for AI research?

The research that fuels the AI revolution has also seen rapid changes. Foremost among them is the maturation of machine learning, stimulated in part by the rise of the digital economy, which both provides and leverages large amounts of data. Other factors include the rise of cloud computing resources and consumer demand for widespread access to services such as speech recognition and navigation support.

Machine learning has been propelled dramatically forward by impressive empirical successes of artificial neural networks, which can now be trained with huge data sets and large-scale computing. This approach has been come to be known as "deep learning." The leap in the performance of information processing algorithms has been accompanied by significant progress in hardware technology for basic operations such as sensing, perception, and object recognition. New platforms and markets for data-driven products, and the economic incentives to find new products and markets, have also stimulated research advances. Now, as it becomes a central force in society, the field of AI is shifting toward building intelligent systems that can collaborate effectively with people, and that are more generally *human-aware*, including creative ways to develop interactive and scalable ways for people to teach robots. These trends drive the currently "hot" areas of AI research into both fundamental methods and application areas:

**Large-scale machine learning** concerns the design of learning algorithms, as well as scaling existing algorithms, to work with extremely large data sets.

**Deep learning**, a class of learning procedures, has facilitated object recognition in images, video labeling, and activity recognition, and is making significant inroads into other areas of perception, such as audio, speech, and natural language processing.

**Reinforcement learning** is a framework that shifts the focus of machine learning from pattern recognition to experience-driven sequential decision-making. It promises to carry AI applications forward toward taking actions in the real world. While largely confined to academia over the past several decades, it is now seeing some practical, real-world successes.

**Robotics** is currently concerned with how to train a robot to interact with the world around it in generalizable and predictable ways, how to facilitate manipulation of objects in interactive environments, and how to interact with people. Advances in robotics will rely on commensurate advances to improve the reliability and generality of computer vision and other forms of machine perception.

**Computer vision** is currently the most prominent form of machine perception. It has been the sub-area of AI most transformed by the rise of deep learning. For the first time, computers are able to perform some vision tasks better than people. Much current research is focused on automatic image and video captioning.

**Natural Language Processing**, often coupled with automatic speech recognition, is quickly becoming a commodity for widely spoken languages with large data sets. Research is now shifting to develop refined and capable systems that are able to interact with people through dialog, not just react to stylized requests. Great strides have also been made in machine translation among different languages, with more real-time person-to-person exchanges on the near horizon.

**Collaborative systems** research investigates models and algorithms to help develop autonomous systems that can work collaboratively with other systems and with humans.

**Crowdsourcing and human computation** research investigates methods to augment computer systems by making automated calls to human expertise to solve problems that computers alone cannot solve well.

**Algorithmic game theory and computational social choice** draw attention to the economic and social computing dimensions of AI, such as how systems can handle potentially misaligned incentives, including self-interested human participants or firms and the automated AI-based agents representing them.

**Internet of Things (IoT)** research is devoted to the idea that a wide array of devices, including appliances, vehicles, buildings, and cameras, can be interconnected to collect and share their abundant sensory information to use for intelligent purposes.

**Neuromorphic computing** is a set of technologies that seek to mimic biological neural networks to improve the hardware efficiency and robustness of computing systems, often replacing an older emphasis on separate modules for input/output, instruction-processing, and memory.

> The field of AI is shifting toward building intelligent systems that can collaborate effectively with people, including creative ways to develop interactive and scalable ways for people to teach robots.

> Misunderstandings about what AI is and is not could fuel opposition to technologies with the potential to benefit everyone. Poorly informed regulation that stifles innovation would be a tragic mistake.

### AI policy, now and in the future

The measure of success for AI applications is the value they create for human lives. In that light, they should be designed to enable people to understand AI systems successfully, participate in their use, and build their trust. Public policies should help ease society's adaptation to AI applications, extend their benefits, and mitigate their inevitable errors and failures. Debate about how AI is deployed, including concerns about how privacy is protected and AI's benefits fairly shared, should be encouraged. Given the speed with which AI technologies are being realized, and concomitant concerns about their implications, the Study Panel recommends that all layers of government acquire technical expertise in AI. Further, research on the fairness, security, privacy, and societal implications of AI systems should be encouraged by removing impediments and increasing private and public spending to support it.

Currently in the United States, at least sixteen separate agencies govern sectors of the economy related to AI technologies. Rapid advances in AI research and, especially, its applications require experts in these sectors to develop new concepts and metaphors for law and policy. Who is responsible when a self-driven car crashes or an intelligent medical device fails? How can AI applications be prevented from promulgating racial discrimination or financial cheating? Who should reap the gains of efficiencies enabled by AI technologies and what protections should be afforded to people whose skills are rendered obsolete? As people integrate AI more broadly and deeply into industrial processes and consumer products, best practices need to be spread, and regulatory regimes adapted.

While the Study Panel does not consider it likely that near-term AI systems will autonomously *choose* to inflict harm on people, it will be possible for people to *use* AI-based systems for harmful as well as helpful purposes. And though AI algorithms may be capable of making less biased decisions than a typical person, it remains a deep technical challenge to ensure that the data that inform AI-based decisions can be kept free from biases that could lead to discrimination based on race, sexual orientation, or other factors.

Faced with the profound changes that AI technologies can produce, pressure for "more" and "tougher" regulation is probably inevitable. Misunderstandings about what AI is and is not could fuel opposition to technologies with the potential to benefit everyone. Inappropriate regulatory activity would be a tragic mistake. Poorly informed regulation that stifles innovation, or relocates it to other jurisdictions, would be counterproductive.[2]

Fortunately, principles that guide successful regulation of current digital technologies provide a starting point. In privacy regulation, broad legal mandates coupled with tough transparency requirements and meaningful enforcement—rather than strict controls—encourage companies to develop processes and professional staff to enforce privacy controls, engage with outside stakeholders, and adapt their practices to technological advances. This in turn supports the development of professional trade associations and standards committees that spread best practices. In AI, too, regulators can strengthen a virtuous cycle of activity involving internal and external accountability, transparency, and professionalization, rather than narrow compliance.

A vigorous and informed debate about how to best steer AI in ways that enrich our lives and our society, while encouraging creativity in the field, is an urgent and vital need. AI technologies could widen existing inequalities of opportunity if access to them—along with the high-powered computation and large-scale data that fuel many of them—is unfairly distributed across society. These technologies will improve the abilities and efficiency of people who have access to them. Policies should be evaluated as to whether they foster democratic values and equitable sharing of AI's benefits, or concentrate power and benefits in the hands of a fortunate few.

---

2    Kate Crawford, "Artificial Intelligence's White Guy Problem," *The New York Times*, June 25, 2016, accessed August 1, 2016, http://www.nytimes.com/2016/06/26/opinion/sunday/artificial-intelligences-white-guy-problem.html.

As this report documents, significant AI-related advances have already had an impact on North American cities over the past fifteen years, and even more substantial developments will occur over the next fifteen. Recent advances are largely due to the growth and analysis of large data sets enabled by the internet, advances in sensory technologies and, more recently, applications of "deep learning." In the coming years, as the public encounters new AI applications in domains such as transportation and healthcare, they must be introduced in ways that build trust and understanding, and respect human and civil rights. While encouraging innovation, policies and processes should address ethical, privacy, and security implications, and should work to ensure that the benefits of AI technologies will be spread broadly and fairly. Doing so will be critical if Artificial Intelligence research and its applications are to exert a positive influence on North American urban life in 2030 and beyond.

# SECTION I: WHAT IS ARTIFICIAL INTELLIGENCE?

*This section describes how researchers and practitioners define "Artificial Intelligence," and the areas of AI research and application that are currently thriving. It proffers definitions of what AI is and is not, and describes some of the currently "hot" areas of AI Research. This section lays the groundwork for Section II, which elaborates on AI's impacts and future in eight domains and Section III, which describes issues related to AI design and public policy and makes recommendations for encouraging AI innovation while protecting democratic values.*

> An accurate and sophisticated picture of AI—one that competes with its popular portrayal—is hampered by the difficulty of pinning down a precise definition of artificial intelligence.

## DEFINING AI

Curiously, the lack of a precise, universally accepted definition of AI probably has helped the field to grow, blossom, and advance at an ever-accelerating pace. Practitioners, researchers, and developers of AI are instead guided by a rough sense of direction and an imperative to "get on with it." Still, a definition remains important and Nils J. Nilsson has provided a useful one:

"Artificial intelligence is that activity devoted to making machines intelligent, and intelligence is that quality that enables an entity to function appropriately and with foresight in its environment."[3]

From this perspective, characterizing AI depends on the credit one is willing to give synthesized software and hardware for functioning "appropriately" and with "foresight." A simple electronic calculator performs calculations much faster than the human brain, and almost never makes a mistake.[4] Is a calculator intelligent? Like Nilsson, the Study Panel takes a broad view that intelligence lies on a multi-dimensional spectrum. According to this view, the difference between an arithmetic calculator and a human brain is not one of kind, but of scale, speed, degree of autonomy, and generality. The same factors can be used to evaluate every other instance of intelligence—speech recognition software, animal brains, cruise-control systems in cars, Go-playing programs, thermostats—and to place them at some appropriate location in the spectrum.

Although our broad interpretation places the calculator within the intelligence spectrum, such simple devices bear little resemblance to today's AI. The frontier of AI has moved far ahead and functions of the calculator are only one among the millions that today's smartphones can perform. AI developers now work on improving, generalizing, and scaling up the intelligence currently found on smartphones.

In fact, the field of AI is a continual endeavor to push forward the frontier of machine intelligence. Ironically, AI suffers the perennial fate of losing claim to its acquisitions, which eventually and inevitably get pulled inside the frontier, a repeating pattern known as the "AI effect" or the "odd paradox"—AI brings a new technology into the common fold, people become accustomed to this technology, it stops being considered AI, and newer technology emerges.[5] The same pattern will continue in the future. AI does not "deliver" a life-changing product as a bolt from the blue. Rather, AI technologies continue to get better in a continual, incremental way.

---

3   Nils J. Nilsson, *The Quest for Artificial Intelligence: A History of Ideas and Achievements* (Cambridge, UK: Cambridge University Press, 2010).
4   Wikimedia Images, accessed August 1, 2016, https://upload.wikimedia.org/wikipedia/commons/b/b6/SHARP_ELSIMATE_EL-W221.jpg.
5   Pamela McCorduck, *Machines Who Think: A Personal Inquiry into the History and Prospects of Artificial Intelligence*, 2nd ed. (Natick, MA: A. K. Peters, Ltd., 2004; San Francisco: W. H. Freeman, 1979), Citations are to the Peters edition.

### The human measure

Notably, the characterization of intelligence as a spectrum grants no special status to the human brain. But to date human intelligence has no match in the biological and artificial worlds for sheer versatility, with the abilities "to reason, achieve goals, understand and generate language, perceive and respond to sensory inputs, prove mathematical theorems, play challenging games, synthesize and summarize information, create art and music, and even write histories."[6]

This makes human intelligence a natural choice for benchmarking the progress of AI. It may even be proposed, as a rule of thumb, that any activity computers are able to perform and people once performed should be counted as an instance of intelligence. But matching any human ability is only a sufficient condition, not a necessary one. There are already many systems that exceed human intelligence, at least in speed, such as scheduling the daily arrivals and departures of thousands of flights in an airport.

AI's long quest—and eventual success—to beat human players at the game of chess offered a high-profile instance for comparing human to machine intelligence. Chess has fascinated people for centuries. When the possibility of building computers became imminent, Alan Turing, who many consider the father of computer science, "mentioned the idea of computers showing intelligence with chess as a paradigm."[7] Without access to powerful computers, "Turing played a game in which he simulated the computer, taking about half an hour per move."

But it was only after a long line of improvements in the sixties and seventies—contributed by groups at Carnegie Mellon, Stanford, MIT, The Institute for Theoretical and Experimental Physics at Moscow, and Northwestern University—that chess-playing programs started gaining proficiency. The final push came through a long-running project at IBM, which culminated with the Deep Blue program beating Garry Kasparov, then the world chess champion, by a score of 3.5-2.5 in 1997. Curiously, no sooner had AI caught up with its elusive target than Deep Blue was portrayed as a collection of "brute force methods" that wasn't "real intelligence."[8] In fact, IBM's subsequent publication about Deep Blue, which gives extensive details about its search and evaluation procedures, doesn't mention the word "intelligent" even once![9] Was Deep Blue intelligent or not? Once again, the frontier had moved.

### An operational definition

AI can also be defined by what AI researchers do. This report views AI primarily as a branch of computer science that studies the properties of intelligence by synthesizing intelligence.[10] Though the advent of AI has depended on the rapid progress of hardware computing resources, the focus here on software reflects a trend in the AI community. More recently, though, progress in building hardware tailored for neural-network-based computing[11] has created a

> Intelligence lies on a multi-dimensional spectrum. According to this view, the difference between an arithmetic calculator and a human brain is not one of kind, but of scale, speed, degree of autonomy, and generality.

---

tighter coupling between hardware and software in advancing AI.

"Intelligence" remains a complex phenomenon whose varied aspects have attracted the attention of several different fields of study, including psychology, economics, neuroscience, biology, engineering, statistics, and linguistics. Naturally, the field of AI has benefited from the progress made by all of these allied fields. For example, the artificial neural network, which has been at the heart of several AI-based solutions[12] [13] was originally inspired by thoughts about the flow of information in biological neurons.[14]

## AI RESEARCH TRENDS

Until the turn of the millennium, AI's appeal lay largely in its promise to deliver, but in the last fifteen years, much of that promise has been redeemed.[15] AI technologies already pervade our lives. As they becomes a central force in society, the field is shifting from simply building systems that are intelligent to building intelligent systems that are human-aware and trustworthy.

Several factors have fueled the AI revolution. Foremost among them is the maturing of machine learning, supported in part by cloud computing resources and wide-spread, web-based data gathering. Machine learning has been propelled dramatically forward by "deep learning," a form of adaptive artificial neural networks trained using a method called backpropagation.[16] This leap in the performance of information processing algorithms has been accompanied by significant progress in hardware technology for basic operations such as sensing, perception, and object recognition. New platforms and markets for data-driven products, and the economic incentives to find new products and markets, have also contributed to the advent of AI-driven technology.

All these trends drive the "hot" areas of research described below. This compilation is meant simply to reflect the areas that, by one metric or another, currently receive greater attention than others. They are not necessarily more important or valuable than other ones. Indeed, some of the currently "hot" areas were less popular in past years, and it is likely that other areas will similarly re-emerge in the future.

**Large-scale machine learning**
Many of the basic problems in machine learning (such as supervised and unsupervised learning) are well-understood. A major focus of current efforts is to scale existing algorithms to work with extremely large data sets. For example, whereas traditional methods could afford to make several passes over the data set, modern ones are designed to make only a single pass; in some cases, only sublinear methods (those that only look at a fraction of the data) can be admitted.

**Deep learning**
The ability to successfully train convolutional neural networks has most benefited the field of computer vision, with applications such as object recognition, video

> Human intelligence has no match in the biological and artificial worlds for sheer versatility, with the abilities "to reason, achieve goals, understand and generate language… create art and music, and even write histories."

---

12  Gerald Tesauro, "Practical Issues in Temporal Difference Learning," *Machine Learning*, no. 8 (1992): 257–77.
13  David Silver, Aja Huang, Chris J. Maddison, Arthur Guez, Laurent Sifre, George van den Driessche, Julian Schrittwieser, Ioannis Antonoglou, Veda Panneershelvam, Marc Lanctot, Sander Dieleman, Dominik Grewe, John Nham, Nal Kalchbrenner, Ilya Sutskever, Timothy Lillicrap, Madeleine Leach, Koray Kavukcuoglu, Thore Graepel, and Demis Hassabis, "Mastering the game of Go with deep neural networks and tree search," *Nature* 529 (2016): 484–489.
14  W. McCulloch and W. Pitts, W., "A logical calculus of the ideas immanent in nervous activity," *Bulletin of Mathematical Biophysics*, 5 (1943): 115–133.
15  Appendix I offers a short history of AI, including a description of some of the traditionally core areas of research, which have shifted over the past six decades.
16  Backpropagation is an abbreviation for "backward propagation of errors," a common method of training artificial neural networks used in conjunction with an optimization method such as gradient descent. The method calculates the gradient of a loss function with respect to all the weights in the network.

labeling, activity recognition, and several variants thereof. Deep learning is also making significant inroads into other areas of perception, such as audio, speech, and natural language processing.

**Reinforcement learning**

Whereas traditional machine learning has mostly focused on pattern mining, reinforcement learning shifts the focus to decision making, and is a technology that will help AI to advance more deeply into the realm of learning about and executing actions in the real world. It has existed for several decades as a framework for experience-driven sequential decision-making, but the methods have not found great success in practice, mainly owing to issues of representation and scaling. However, the advent of deep learning has provided reinforcement learning with a "shot in the arm." The recent success of AlphaGo, a computer program developed by Google Deepmind that beat the human Go champion in a five-game match, was due in large part to reinforcement learning. AlphaGo was trained by initializing an automated agent with a human expert database, but was subsequently refined by playing a large number of games against itself and applying reinforcement learning.

**Robotics**

Robotic navigation, at least in static environments, is largely solved. Current efforts consider how to train a robot to interact with the world around it in generalizable and predictable ways. A natural requirement that arises in interactive environments is *manipulation*, another topic of current interest. The deep learning revolution is only beginning to influence robotics, in large part because it is far more difficult to acquire the large labeled data sets that have driven other learning-based areas of AI. Reinforcement learning (see above), which obviates the requirement of labeled data, may help bridge this gap but requires systems to be able to safely explore a policy space without committing errors that harm the system itself or others. Advances in reliable machine perception, including computer vision, force, and tactile perception, much of which will be driven by machine learning, will continue to be key enablers to advancing the capabilities of robotics.

**Computer vision**

Computer vision is currently the most prominent form of machine perception. It has been the sub-area of AI most transformed by the rise of deep learning. Until just a few years ago, support vector machines were the method of choice for most visual classification tasks. But the confluence of large-scale computing, especially on GPUs, the availability of large datasets, especially via the internet, and refinements of neural network algorithms has led to dramatic improvements in performance on benchmark tasks (e.g., classification on ImageNet[17]). For the first time, computers are able to perform some (narrowly defined) visual classification tasks better than people. Much current research is focused on automatic image and video captioning.

**Natural Language Processing**

Often coupled with automatic speech recognition, Natural Language Processing is another very active area of machine perception. It is quickly becoming a commodity for mainstream languages with large data sets. Google announced that 20% of current mobile queries are done by voice,[18] and recent demonstrations have proven the possibility of real-time translation. Research is now shifting towards developing refined and capable systems that are able to interact with people through dialog, not just react to stylized requests.

> AI technologies already pervade our lives. As they become a central force in society, the field is shifting from simply building systems that are intelligent to building intelligent systems that are human-aware and trustworthy.

---

17   ImageNet, Stanford Vision Lab, Stanford University, Princeton University, 2016, accessed August 1, 2016, www.image-net.org/.
18   Greg Sterling, "Google says 20% of mobile queries are voice searches," *Search Engine Land*, May 18, 2016, accessed August 1, 2016, http://searchengineland.com/google-reveals-20-percent-queries-voice-queries-249917.

> Natural Language Processing is a very active area of machine perception. Research is now shifting towards developing systems that are able to interact with people through dialog, not just react to stylized requests.

**Collaborative systems**
Research on collaborative systems investigates models and algorithms to help develop autonomous systems that can work collaboratively with other systems and with humans. This research relies on developing formal models of collaboration, and studies the capabilities needed for systems to become effective partners. There is growing interest in applications that can utilize the complementary strengths of humans and machines—for humans to help AI systems to overcome their limitations, and for agents to augment human abilities and activities.

**Crowdsourcing and human computation**
Since human abilities are superior to automated methods for accomplishing many tasks, research on crowdsourcing and human computation investigates methods to augment computer systems by utilizing human intelligence to solve problems that computers alone cannot solve well. Introduced only about fifteen years ago, this research now has an established presence in AI. The best-known example of crowdsourcing is Wikipedia, a knowledge repository that is maintained and updated by netizens and that far exceeds traditionally-compiled information sources, such as encyclopedias and dictionaries, in scale and depth. Crowdsourcing focuses on devising innovative ways to harness human intelligence. Citizen science platforms energize volunteers to solve scientific problems, while paid crowdsourcing platforms such as Amazon Mechanical Turk provide automated access to human intelligence on demand. Work in this area has facilitated advances in other subfields of AI, including computer vision and NLP, by enabling large amounts of labeled training data and/or human interaction data to be collected in a short amount of time. Current research efforts explore ideal divisions of tasks between humans and machines based on their differing capabilities and costs.

**Algorithmic game theory and computational social choice**
New attention is being drawn to the economic and social computing dimensions of AI, including incentive structures. Distributed AI and multi-agent systems have been studied since the early 1980s, gained prominence starting in the late 1990s, and were accelerated by the internet. A natural requirement is that systems handle potentially misaligned incentives, including self-interested human participants or firms, as well as automated AI-based agents representing them. Topics receiving attention include computational mechanism design (an economic theory of incentive design, seeking incentive-compatible systems where inputs are truthfully reported), computational social choice (a theory for how to aggregate rank orders on alternatives), incentive aligned information elicitation (prediction markets, scoring rules, peer prediction) and algorithmic game theory (the equilibria of markets, network games, and parlor games such as Poker—a game where significant advances have been made in recent years through abstraction techniques and no-regret learning).

**Internet of Things (IoT)**
A growing body of research is devoted to the idea that a wide array of devices can be interconnected to collect and share their sensory information. Such devices can include appliances, vehicles, buildings, cameras, and other things. While it's a matter of technology and wireless networking to connect the devices, AI can process and use the resulting huge amounts of data for intelligent and useful purposes. Currently, these devices use a bewildering array of incompatible communication protocols. AI could help tame this Tower of Babel.

**Neuromorphic Computing**
Traditional computers implement the von Neumann model of computing, which separates the modules for input/output, instruction-processing, and memory. With the success of deep neural networks on a wide array of tasks, manufacturers are actively pursuing alternative models of computing—especially those that are inspired by what is known about biological neural networks—with the aim of improving the hardware efficiency and robustness of computing systems. At the moment, such "neuromorphic" computers have not yet clearly demonstrated big wins, and are just beginning to become commercially viable. But it is possible that they will become commonplace (even if only as additions to their von Neumann cousins) in the near future. Deep neural networks have already created a splash in the application landscape. A larger wave may hit when these networks can be trained and executed on dedicated neuromorphic hardware, as opposed to simulated on standard von Neumann architectures, as they are today.

### Overall trends and the future of AI research

The resounding success of the data-driven paradigm has displaced the traditional paradigms of AI. Procedures such as theorem proving and logic-based knowledge representation and reasoning are receiving reduced attention, in part because of the ongoing challenge of connecting with real-world groundings. Planning, which was a mainstay of AI research in the seventies and eighties, has also received less attention of late due in part to its strong reliance on modeling assumptions that are hard to satisfy in realistic applications. Model-based approaches—such as physics-based approaches to vision and traditional control and mapping in robotics—have by and large given way to data-driven approaches that close the loop with sensing the results of actions in the task at hand. Bayesian reasoning and graphical models, which were very popular even quite recently, also appear to be going out of favor, having been drowned by the deluge of data and the remarkable success of deep learning.

Over the next fifteen years, the Study Panel expects an increasing focus on developing systems that are human-aware, meaning that they specifically model, and are specifically designed for, the characteristics of the people with whom they are meant to interact. There is a lot of interest in trying to find new, creative ways to develop interactive and scalable ways to teach robots. Also, IoT-type systems—devices and the cloud—are becoming increasingly popular, as is thinking about social and economic dimensions of AI. In the coming years, new perception/object recognition capabilities and robotic platforms that are human-safe will grow, as will data-driven products and their markets.

The Study Panel also expects a reemergence of some of the traditional forms of AI as practitioners come to realize the inevitable limitations of purely end-to-end deep learning approaches. We encourage young researchers not to reinvent the wheel, but rather to maintain an awareness of the significant progress in many areas of AI during the first fifty years of the field, and in related fields such as control theory, cognitive science, and psychology.

> **A growing body of research is devoted to the idea that a wide array of devices can be interconnected to collect and share their sensory information. Such devices can include appliances, vehicles, buildings, cameras, and other things.**

# SECTION II: AI BY DOMAIN

*Though different instances of AI research and practice share common technologies, such as machine learning, they also vary considerably in different sectors of the economy and society. We call these sectors "domains," and in this section describe the different states of AI research and implementation, as well as impacts and distinct challenges, in eight of them: transportation; home/service robotics; healthcare; education; low-resource communities; public safety and security; employment and workplace; and entertainment. Based on these analyses, we also predict trends in a typical North American city over the next fifteen years. Contrary to AI's typical depiction in popular culture, we seek to offer a balanced overview of the ways in which AI is already beginning to transform everyday life, and how those transformations are likely to grow by the year 2030.*

> Autonomous transportation will soon be commonplace and, as most people's first experience with physically embodied AI systems, will strongly influence the public's perception of AI.

## TRANSPORTATION

Transportation is likely to be one of the first domains in which the general public will be asked to trust the reliability and safety of an AI system for a critical task. Autonomous transportation will soon be commonplace and, as most people's first experience with physically embodied AI systems, will strongly influence the public's perception of AI. Once the physical hardware is made sufficiently safe and robust, its introduction to daily life may happen so suddenly as to surprise the public, which will require time to adjust. As cars will become better drivers than people, city-dwellers will own fewer cars, live further from work, and spend time differently, leading to an entirely new urban organization. Further, in the typical North American city in 2030, changes won't be limited to cars and trucks, but are likely to include flying vehicles and personal robots, and will raise social, ethical and policy issues.

A few key technologies have already catalyzed the widespread adoption of AI in transportation. Compared to 2000, the scale and diversity of data about personal and population-level transportation available today—enabled by the adoption of smartphones and decreased costs and improved accuracies for variety of sensors—is astounding. Without the availability of this data and connectivity, applications such as real-time sensing and prediction of traffic, route calculations, peer-to-peer ridesharing and self-driving cars would not be possible.

### Smarter cars

GPS was introduced to personal vehicles in 2001 with in-car navigation devices and has since become a fundamental part of the transportation infrastructure.[19] GPS assists drivers while providing large-scale information to technology companies and cities about transportation patterns. Widespread adoption of smartphones with GPS technology further increased connectivity and the amount of location data shared by individuals.

Current vehicles are also equipped with a wide range of sensing capabilities. An average automobile in the US is predicted to have seventy sensors including gyroscopes, accelerometers, ambient light sensors, and moisture sensors.[20] Sensors are not new to vehicles. Automobiles built before 2000 had sensors for the internal state of the vehicle such as its speed, acceleration, and wheel position.[21]

---

19  Mark Sullivan, "A brief history of GPS," *PCWorld*, August 9, 2012, accessed August 1, 2016, http://www.pcworld.com/article/2000276/a-brief-history-of-gps.html.
20  William J. Fleming, "New Automotive Sensors - A Review," *IEEE Sensors Journal* 8, *no 11*, (2008): 1900-1921.
21  Jean Jacques Meneu, ed., "Automotive Sensors: Now and in the Future," *Arrow*, September 24, 2015, accessed August 1, 2016, https://www.arrow.com/en/research-and-events/articles/automotive-sensors-now-and-in-the-future.

They already had a number of functionalities that combined real-time sensing with perception and decision-making such as Anti-lock Braking Systems (ABS), airbag control, Traction Control Systems (TCS), and Electronic Stability Control (ESC).[22] Automated capabilities have been introduced into commercial cars gradually since 2003 as summarized in the following table.

| Context | Automated Functionality | Release Date |
| --- | --- | --- |
| Parking | Intelligent Parking Assist System | Since 2003[23] |
| Parking | Summon | Since 2016[24] |
| Arterial & Highway | Lane departure system | Since 2004 in North America[25] |
| Arterial & Highway | Adaptive cruise control | Since 2005 in North America[26] |
| Highway | Blind spot monitoring | 2007[27] |
| Highway | Lane changing | 2015[28] |

These functionalities assist drivers or completely take over well-defined activities for increased safety and comfort. Current cars can park themselves, perform adaptive cruise control on highways, steer themselves during stop-and-go traffic, and alert drivers about objects in blind spots during lane changes. Vision and radar technology were leveraged to develop pre-collision systems that let cars autonomously brake when risk of a collision is detected. Deep learning also has been applied to improve automobiles' capacity to detect objects in the environment and recognize sound.[29]

### Self-driving vehicles

Since the 1930s, science fiction writers dreamed of a future with self-driving cars, and building them has been a challenge for the AI community since the 1960s. By the 2000s, the dream of autonomous vehicles became a reality in the sea and sky, and even on Mars, but self-driving cars existed only as research prototypes in labs. Driving in a city was considered to be a problem too complex for automation due to factors like pedestrians, heavy traffic, and the many unexpected events that can happen outside of the car's control. Although the technological components required to

**As cars will become better drivers than people, city-dwellers will own fewer cars, live further from work, and spend time differently, leading to an entirely new urban organization.**

---

22  Carl Liersch, "Vehicle Technology Timeline: From Automated to Driverless," Robert Bosch (Australia) Pty. Ltd., 2014, accessed August 1, 2016, http://dpti.sa.gov.au/__data/assets/pdf_file/0009/246807/Carl_Liersch_Presentation.pdf.
23  "Intelligent Parking Assist System," *Wikipedia*, last modified July 26, 2016, accessed August 1, 2016, https://en.wikipedia.org/wiki/Intelligent_Parking_Assist_System.
24  The Tesla Motors Team, "Summon Your Tesla from Your Phone," Tesla, January 10, 2016, accessed August 1, 2016, https://www.teslamotors.com/blog/summon-your-tesla-your-phone.
25  Lane departure warning system," *Wikipedia*, last modified July 24, 2016, accessed August 1, 2016, https://en.wikipedia.org/wiki/Lane_departure_warning_system.
26  "Autonomous cruise control system," *Wikipedia*, last modified July 30, 2016, accessed August 1, 2016, https://en.wikipedia.org/wiki/Autonomous_cruise_control_system.
27  "Blind spot monitor," *Wikipedia*, last modified April 20, 2016, accessed August 1, 2016, https://en.wikipedia.org/wiki/Blind_spot_monitor.
28  Dana Hull, "Tesla Starts Rolling Out Autopilot Features," *Boomberg Technology*, October 14, 2015, accessed August 1, 2016, http://www.bloomberg.com/news/articles/2015-10-14/tesla-software-upgrade-adds-automated-lane-changing-to-model-s.
29  Aaron Tilley, "New Qualcomm Chip Brings Deep Learning To Cars," *Forbes*, January 5, 2016, accessed August 1, 2016, http://www.forbes.com/sites/aarontilley/2016/01/05/along-with-nvidia-new-qualcomm-chip-brings-deep-learning-to-cars/#4cb4e9235357.

> We will see self-driving and remotely controlled delivery vehicles, flying vehicles, and trucks. Peer-to-peer transportation services such as ridesharing are also likely to utilize self-driving vehicles.

make such autonomous driving possible were available in 2000—and indeed some autonomous car prototypes existed[30][31][32]—few predicted that mainstream companies would be developing and deploying autonomous cars by 2015. During the first Defense Advanced Research Projects Agency (DARPA) "grand challenge" on autonomous driving in 2004, research teams failed to complete the challenge in a limited desert setting.

But in eight short years, from 2004-2012, speedy and surprising progress occurred in both academia and industry. Advances in sensing technology and machine learning for perception tasks has sped progress and, as a result, Google's autonomous vehicles and Tesla's semi-autonomous cars are driving on city streets today. Google's self-driving cars, which have logged more than 1,500,000 miles (300,000 miles without an accident),[33] are completely autonomous—no human input needed. Tesla has widely released self-driving capability to existing cars with a software update.[34] Their cars are semi-autonomous, with human drivers expected to stay engaged and take over if they detect a potential problem. It is not yet clear whether this semi-autonomous approach is sustainable, since as people become more confident in the cars' capabilities, they are likely to pay less attention to the road, and become less reliable when they are most needed. The first traffic fatality involving an autonomous car, which occurred in June of 2016, brought this question into sharper focus.[35]

In the near future, sensing algorithms will achieve super-human performance for capabilities required for driving. Automated perception, including vision, is already near or at human-level performance for well-defined tasks such as recognition and tracking. Advances in perception will be followed by algorithmic improvements in higher level reasoning capabilities such as planning. A recent report predicts self-driving cars to be widely adopted by 2020.[36] And the adoption of self-driving capabilities won't be limited to personal transportation. We will see self-driving and remotely controlled delivery vehicles, flying vehicles, and trucks. Peer-to-peer transportation services such as ridesharing are also likely to utilize self-driving vehicles. Beyond self-driving cars, advances in robotics will facilitate the creation and adoption of other types of autonomous vehicles, including robots and drones.

It is not yet clear how much better self-driving cars need to become to encourage broad acceptance. The collaboration required in semi-self-driving cars and its implications for the cognitive load of human drivers is not well understood. But if future self-driving cars are adopted with the predicted speed, and they exceed human-level performance in driving, other significant societal changes will follow. Self-driving cars will eliminate one of the biggest causes of accidental death and injury in United States, and lengthen people's life expectancy. On average, a commuter in US spends twenty-five minutes driving each way.[37] With self-driving car technology, people will have more time to work or entertain themselves during their commutes. And the increased comfort and decreased cognitive load with self-driving cars and shared transportation may affect where people choose to live. The reduced need for parking may affect the way cities and public spaces are designed. Self-driving cars may also serve to increase the freedom and mobility of different subgroups of the population, including youth, elderly and disabled.

---

30  "Navlab," *Wikipedia*, last updated June 4, 2016, accessed August 1, 2016, https://en.wikipedia.org/wiki/Navlab.
31  "Navlab: The Carnegie Mellon University Navigation Laboratory," Carnegie Mellon University, accessed August 1, 2016, http://www.cs.cmu.edu/afs/cs/project/alv/www/.
32  "Eureka Prometheus Project," *Wikipedia*, last modified February 12, 2016, accessed August 1, 2016, https://en.wikipedia.org/wiki/Eureka_Prometheus_Project.
33  "Google Self-Driving Car Project," Google, accessed August 1, 2016, https://www.google.com/selfdrivingcar/. 33    Molly McHugh, "Tesla's Cars Now Drive Themselves, Kinda," *Wired*, October 14, 2015, accessed August 1, 2016, http://www.wired.com/2015/10/tesla-self-driving-over-air-update-live/.
34  Molly McHugh, "Tesla's Cars Now Drive Themselves, Kinda," *Wired*, October 14, 2015, accessed August 1, 2016, http://www.wired.com/2015/10/tesla-self-driving-over-air-update-live/.
35  Anjali Singhvi and Karl Russell, "Inside the Self-Driving Tesla Fatal Accident," *The New York Times*, Last updated July 12, 2016, accessed August 1, 2016, http://www.nytimes.com/interactive/2016/07/01/business/inside-tesla-accident.html.
36  John Greenough, "10 million self-driving cars will be on the road by 2020," *Business Insider*, June 15, 2016, accessed August 1, 2016, http://www.businessinsider.com/report-10-million-self-driving-cars-will-be-on-the-road-by-2020-2015-5-6.

Self-driving cars and peer-to-peer transportation services may eliminate the need to own a vehicle. The effect on total car use is hard to predict. Trips of empty vehicles and people's increased willingness to travel may lead to more total miles driven. Alternatively, shared autonomous vehicles—people using cars as a service rather than owning their own—may reduce total miles, especially if combined with well-constructed incentives, such as tolls or discounts, to spread out travel demand, share trips, and reduce congestion. The availability of shared transportation may displace the need for public transportation—or public transportation may change form towards personal rapid transit, already available in four cities,[38] which uses small capacity vehicles to transport people on demand and point-to-point between many stations.[39]

As autonomous vehicles become more widespread, questions will arise over their security, including how to ensure that technologies are safe and properly tested under different road conditions prior to their release. Autonomous vehicles and the connected transportation infrastructure will create a new venue for hackers to exploit vulnerabilities to attack. Ethical questions are also involved in programming cars to act in situations in which human injury or death is inevitable, especially when there are split-second choices to be made about whom to put at risk. The legal systems in most states in the US do not have rules covering self-driving cars. As of 2016, four states in the US (Nevada, Florida, California, and Michigan), Ontario in Canada, the United Kingdom, France, and Switzerland have passed rules for the testing of self-driving cars on public roads. Even these laws do not address issues about responsibility and assignment of blame for an accident for self-driving and semi-self-driving cars.[40]

### Transportation planning

By 2005, cities had started investing in the transportation infrastructure to develop sensing capabilities for vehicle and pedestrian traffic.[41] The sensors currently used include inductive loops, video cameras, remote traffic microwave sensors, radars, and GPS.[42] For example, in 2013 New York started using a combination of microwave sensors, a network of cameras, and pass readers to detect vehicle traffic in the city.[43]

> Shared transportation may displace the need for public transportation— or public transportation may change form towards personal rapid transit that uses small capacity vehicles to transport people on demand.

---

37    Brian McKenzie and Melanie Rapino, "Commuting in the United States: 2009," *American Community Survey Reports*, United States Census Bureau, September 2011, accessed August 1, 2016, https://www.census.gov/prod/2011pubs/acs-15.pdf.
38    Morgantown, West Virginia; Masdar City, UAE; London, England; and Suncheon, South Korea.
39    "Personal rapid transit," *Wikipedia*, Last modified July 18, 2016, accessed August 1, 2016, https://en.wikipedia.org/wiki/Personal_rapid_transit.
40    Patrick Lin, "The Ethics of Autonomous Cars," *The Atlantic*, October 8, 2013, accessed August 1, 2016, http://www.theatlantic.com/technology/archive/2013/10/the-ethics-of-autonomous-cars/280360/.
41    Steve Lohr, "Bringing Efficiency to the Infrastructure," *The New York Times*, April 29, 2009, accessed August 1, 2016, http://www.nytimes.com/2009/04/30/business/energy-environment/30smart.html.
42    "Intelligent transportation system," *Wikipedia*, last modified July 28, 2016, accessed August 1, 2016, https://en.wikipedia.org/wiki/Intelligent_transportation_system.
43    Access Science Editors, "Active traffic management: adaptive traffic signal control," *Access Science*, 2014, accessed August 1, 2016, http://www.accessscience.com/content/active-traffic-management-adaptive-traffic-signal-control/BR0106141.

> Ethical questions arise when programming cars to act in situations in which human injury or death is inevitable, especially when there are split-second choices to be made about whom to put at risk.

Cities use AI methods to optimize services in several ways, such as bus and subway schedules, and tracking traffic conditions to dynamically adjust speed limits or apply smart pricing on highways, bridges, and HOV lanes.[44][45][46] Using sensors and cameras in the road network, they can also optimize traffic light timing for improving traffic flow and to help with automated enforcement.[47][48] These dynamic strategies are aimed at better utilizing the limited resources in the transportation network, and are made possible by the availability of data and the widespread connectivity of individuals.

Before the 2000s, transportation planners were forced to rely on static pricing strategies tied to particular days or times of day, to manage demand. As dynamic pricing strategies are adopted, this raises new issues concerning the fair distribution of public goods, since market conditions in high-demand situations may make services unavailable to segments of the public.

The availability of large-scale data has also made transportation an ideal domain for machine learning applications. Since 2006, applications such as Mapquest, Google Maps, and Bing Maps have been widely used by the public for routing trips, using public transportation, receiving real-time information and predictions about traffic conditions,[49][50] and finding services around a location.[51][52] Optimal search algorithms have been applied to the routing of vehicles and pedestrians to a given destination (i.e.[53][54]).

Despite these advances, the widespread application of sensing and optimization techniques to city infrastructure has been slower than the application of these techniques to individual vehicles or people. Although individual cities have implemented sensing and optimization applications, as yet there is no standardization of the sensing infrastructure and AI techniques used. Infrastructure costs, differing priorities among cities, and the high coordination costs among the parties involved have slowed adoption, as have public concerns over privacy related to sensing. Still,

AI is likely to have an increasing impact on city infrastructure. Accurate predictive models of individuals' movements, their preferences, and their goals are likely to emerge with the greater availability of data. The ethical issues regarding such an emergence are discussed in Section III of this report.

The United States Department of Transportation released a call for proposals in 2016 asking medium-size cities to imagine smart city infrastructure for transportation.[55] This initiative plans to award forty million dollars to a city to demonstrate how technology and data can be used to reimagine the movement of people as well as goods.

One vision is a network of connected vehicles that can reach a high level of safety in driving with car-to-car communication.[56] If this vision becomes reality, we expect advances in multi-agent coordination, collaboration, and planning will have a significant impact on future cars and play a role in making the transportation system more reliable and efficient. Robots are also likely to take part in transportation by carrying individuals and packages (c.f., Segway robot). For transportation of goods, interest in drones has increased, and Amazon is currently testing a delivery system using them,[57] although questions remain about the appropriate safety rules and regulations.

The increased sensing capabilities, adoption of drones, and the connected transportation infrastructure will also raise concerns about the privacy of individuals and the safety of private data. In coming years, these and related transportation issues will need to be addressed either by preemptive action on the part of industry or within the legal framework. As noted in the Section III policy discussion, how well this is done will affect the pace and scope of AI-related advances in the transportation sector.

> Our Study Panel doesn't expect drones that can fly, swim, and drive, or flying quadcoptors to become a common means of transportation by 2030 (although prototypes exist today).

## On-demand transportation

On-demand transportation services such as Uber and Lyft have emerged as another pivotal application of sensing, connectivity, and AI,[58] with algorithms for matching drivers to passengers by location and suitability (reputation modeling).[59][60]

Through dynamic pricing, these services ration access by willingness-to-pay, with dynamic pricing also encouraging an increase in the supply of drivers, and have become a popular method for transportation in cities. With their rapid advance have come multiple policy and legal issues, such as competition with existing taxi services and concerns about lack of regulation and safety. On-demand transportation services seem likely to be a major force towards self-driving cars.

Carpooling and ridesharing have long been seen as a promising approach to decrease traffic congestion and better utilize personal transportation resources. Services such as Zimride and Nuride bring together people sharing similar routes for a joint trip. But this approach to carpooling has failed to gain traction on a large scale.

---

55  "U.S. Department of Transportation Launches Smart City Challenge to Create a City of the Future," Transportation.gov, U.S. Department of Transportation, December 7, 2015, accessed August 1, 2016, https://www.transportation.gov/briefing-room/us-department-transportation-launches-smart-city-challenge-create-city-future.
56  Will Knight, "Car-to-Car Communication: A simple wireless technology promises to make driving much safer.," *MIT Technology Review*, accessed August 1, 2016, https://www.technologyreview.com/s/534981/car-to-car-communication/.
57  "Amazon Prime Air," Amazon, accessed August 1, 2016, http://www.amazon.com/b?node=8037720011.
58  Jared Meyer, "Uber and Lyft are changing the way Americans move about their country," *National Review*, June 7, 2016, accessed August 1, 2016, http://www.nationalreview.com/article/436263/uber-lyft-ride-sharing-services-sharing-economy-are-future.
59  Alexander Howard, "How Digital Platforms Like LinkedIn, Uber And TaskRabbit Are Changing The On-Demand Economy," *The Huffington Post*, July 14, 2015, accessed August 1, 2016, http://www.huffingtonpost.com/entry/online-talent-platforms_us_55a03545e4b0b8145f72ccf6.
60  "Announcing UberPool," Uber Newsroom, August 5, 2014, accessed August 1, 2016, https://newsroom.uber.com/announcing-uberpool/.

> Over the next fifteen years, coincident advances in mechanical and AI technologies promise to increase the safe and reliable use and utility of home robots in a typical North American city.

### Interacting with people

For decades, people have imagined wildly different, futuristic-looking transportation vehicles. Although future cars will be smarter and drones will be available widely, it is unlikely that by 2030 we will have widely adopted transportation vehicles that look and function differently than the ones we have today. Our Study Panel doesn't expect drones that can fly, swim, and drive, or flying quadcoptors to become a common means of transportation in this time horizon (although prototypes exist today).

We do expect humans to become partners to self-driving cars and drones in their training, execution, and evaluation. This partnering will happen both when humans are co-located with machines and also virtually. We predict advances in algorithms to facilitate machine learning from human input. We also expect models and algorithms for modeling of human attention, and to support communication and coordination between humans and machine. This is an integral part of the development of future vehicles.

## HOME/SERVICE ROBOTS

Robots have entered people's homes in the past fifteen years. Disappointingly slow growth in the diversity of applications has occurred simultaneously with increasingly sophisticated AI deployed on existing applications. AI advances are often inspired by mechanical innovations, which in turn prompt new AI techniques to be introduced.

Over the next fifteen years, coincident advances in mechanical and AI technologies promise to increase the safe and reliable use and utility of home robots in a typical North American city. Special purpose robots will deliver packages, clean offices, and enhance security, but technical constraints and the high costs of reliable mechanical devices will continue to limit commercial opportunities to narrowly defined applications for the foreseeable future. As with self-driving cars and other new transportation machines, the difficulty of creating reliable, market-ready hardware is not to be underestimated.

### Vacuum cleaners

In 2001, after many years of development, the Electrolux Trilobite, a vacuum cleaning robot, became the first commercial home robot. It had a simple control system to do obstacle avoidance, and some navigation. A year later, iRobot introduced Roomba, which was a tenth the price of the Trilobite and, with only 512 bytes of RAM, ran a behavior based controller. The most intelligent thing it did was to avoid falling down stairs. Since then, sixteen million Roombas have been deployed all over the world and several other competing brands now exist.

As the processing power and RAM capacity of low cost embedded processors improved from its dismal state in the year 2000, the AI capabilities of these robots also improved dramatically. Simple navigation, self-charging, and actions for dealing with full dust bins were added, followed by ability to deal with electrical cords and rug tassels, enabled by a combination of mechanical improvements and sensor based perception. More recently, the addition of full VSLAM (Visual Simultaneous Location and Mapping)— an AI technology that had been around for twenty years— has enabled the robots to build a complete 3D world model of a house as they clean, and become more efficient in their cleaning coverage.

Early expectations that many new applications would be found for home robots have not materialized. Robot vacuum cleaners are restricted to localized flat areas, while real homes have lots of single steps, and often staircases; there has been very little research on robot mobility inside real homes. Hardware platforms remain challenging to build, and there are few applications that people want enough to buy. Perceptual algorithms for functions such as image labeling, and 3D object recognition, while common at AI conferences, are still only a few years into development as products.

**Home robots 2030**

Despite the slow growth to date of robots in the home, there are signs that this will change in the next fifteen years. Corporations such as Amazon Robotics and Uber are developing large economies of scale using various aggregation technologies. Also:

System in Module (SiM), with a lot of System on Chip (SoC) subsystems, are now being pushed out the door by phone-chip makers (Qualcomm's SnapDragon, Samsung's Artik, etc.). These are better than supercomputers of less than ten years ago with eight or more sixty-four-bit cores, and specialized silicon for cryptography, camera drivers, additional DSPs, and hard silicon for certain perceptual algorithms. This means that low cost devices will be able to support much more onboard AI than we have been able to consider over the last fifteen years.

Cloud ("someone else's computer") is going to enable more rapid release of new software on home robots, and more sharing of data sets gathered in many different homes, which will in turn feed cloud-based machine learning, and then power improvements to already deployed robots.

The great advances in speech understanding and image labeling enabled by deep learning will enhance robots' interactions with people in their homes.

Low cost 3D sensors, driven by gaming platforms, have fueled work on 3D perception algorithms by thousands of researchers worldwide, which will speed the development and adoption of home and service robots.

In the past three years, low cost and safe robot arms have been introduced to hundreds of research labs around the world, sparking a new class of research on manipulation that will eventually be applicable in the home, perhaps around 2025. More than half a dozen startups around the world are developing AI-based robots for the home, for now concentrating mainly on social interaction. New ethics and privacy issues may surface as a result.

## HEALTHCARE

For AI technologies, healthcare has long been viewed as a promising domain. AI-based applications could improve health outcomes and quality of life for millions of people in the coming years—but only if they gain the trust of doctors, nurses, and patients, and if policy, regulatory, and commercial obstacles are removed. Prime applications include clinical decision support, patient monitoring and coaching, automated devices to assist in surgery or patient care, and management of healthcare systems. Recent successes, such as mining social media to infer possible health risks, machine learning to predict patients at risk, and robotics to support surgery, have expanded a sense of possibility for AI in healthcare. Improvements in methods for interacting with medical professionals and patients will be a critical challenge.

As in other domains, data is a key enabler. There has been an immense forward leap in collecting useful data from personal monitoring devices and mobile apps, from electronic health records (EHR) in clinical settings and, to a lesser extent, from robots designed to assist with medical procedures and hospital operations. But using this data to enable more finely-grained diagnostics and treatments for both individual patients and patient populations has proved difficult. Research and deployment have been slowed by outdated regulations and incentive structures. Poor human-computer interaction methods and the inherent difficulties and risks of implementing technologies in such a large and complex system have slowed realization of AI's

> **Special purpose robots will deliver packages, clean offices, and enhance security, but technical constraints and high costs will continue to limit commercial opportunities for the foreseeable future.**

> AI-based applications could improve health outcomes and quality of life for millions of people in the coming years—but only if they gain the trust of doctors, nurses, and patients.

promise in healthcare.[61] The reduction or removal of these obstacles, combined with innovations still on the horizon, have the potential to significantly improve health outcomes and quality of life for millions of people in the coming years.

## The clinical setting

For decades, the vision of an AI-powered clinician's assistant has been a near cliché. Although there have been successful pilots of AI-related technology in healthcare,[62] the current healthcare delivery system unfortunately remains structurally ill-suited to absorb and deploy rapid advances. Incentives provided by the Affordable Care Act have accelerated the penetration of electronic health records (EHRs) into clinical practice, but implementation has been poor, eroding clinicians' confidence in their usefulness. A small group of companies control the EHR market, and user interfaces are widely considered substandard, including annoying pop-ups that physicians routinely dismiss. The promise of new analytics using data from EHRs, including AI, remains largely unrealized due to these and other regulatory and structural barriers.

Looking ahead to the next fifteen years, AI advances, if coupled with sufficient data and well-targeted systems, promise to change the cognitive tasks assigned to human clinicians. Physicians now routinely solicit verbal descriptions of symptoms from presenting patients and, in their heads, correlate patterns against the clinical presentation of known diseases. With automated assistance, the physician could instead supervise this process, applying her or his experience and intuition to guide the input process and to evaluate the output of the machine intelligence. The literal "hands-on" experience of the physician will remain critical. A significant challenge is to optimally integrate the human dimensions of care with automated reasoning processes.

To achieve future advances, clinicians must be involved and engaged at the outset to ensure that systems are well-engineered and trusted. Already, a new generation of more tech savvy physicians routinely utilize specialized apps on mobile devices. At the same time, workloads on primary care clinicians have increased to the point that they are grateful for help from any quarter. Thus, the opportunity to exploit new learning methods, to create structured patterns of inference by mining the scientific literature automatically, and to create true cognitive assistants by supporting free-form dialogue, has never been greater. Provided these advances are not stymied by regulatory, legal, and social barriers, immense improvements to the value of healthcare are within our grasp.

## Healthcare analytics

At the population level, AI's ability to mine outcomes from millions of patient clinical records promises to enable finer-grained, more personalized diagnosis and treatment. Automated discovery of genotype-phenotype connections will also become possible as full, once-in-a-lifetime genome sequencing becomes routine for each patient.

A related (and perhaps earlier) capability will be to find "patients like mine" as a way to inform treatment decisions based on analysis of a similar cohort. Traditional and non-traditional healthcare data, augmented by social platforms, may lead to the emergence of self-defined subpopulations, each managed by a surrounding ecosystem of healthcare providers augmented with automated recommendation and monitoring systems. These developments have the potential to radically transform healthcare delivery as medical procedures and lifetime clinical records for hundreds of millions of individuals become available. Similarly, the automated capture of personal environmental data from wearable devices will expand personalized medicine. These activities are becoming more commercially viable as vendors discover ways to engage large populations (e.g. ShareCare)[63] and then to create population-scale data that can be mined to produce individualized analytics and recommendations.

---

61   LeighAnne Olsen, Dara Aisner, and J. Michael McGinnis, eds., "Institute of Medicine (US) Roundtable on Evidence-Based Medicine," *The Learning Healthcare System: Workshop Summary*. (Washington (DC): *National Academies Press* (US); 2007), accessed August 1, 2016, http://www.ncbi.nlm.nih.gov/books/NBK53500/.

62   Katherine E. Henry, David N. Hager, Peter J. Pronovost, and Suchi Saria, "A Targeted Real-time Early Warning Score (TREWScore) for Septic Shock," *Science Translational Medicine 7*, (299), 299ra122.

Unfortunately, the FDA has been slow to approve innovative diagnostic software, and there are many remaining barriers to rapid innovation. HIPAA (Health Insurance Portability and Accountability Act) requirements for protecting patient privacy create legal barriers to the flow of patient data to applications that could utilize AI technologies. Unanticipated negative effects of approved drugs could show up routinely, sooner, and more rigorously than they do today, but mobile apps that analyze drug interactions may be blocked from pulling the necessary information from patient records. More generally, AI research and innovation in healthcare are hampered by the lack of widely accepted methods and standards for privacy protection. The FDA has been slow to approve innovative software, in part due to an unclear understanding of the cost/benefit tradeoffs of these systems. If regulators (principally the FDA) recognize that effective post-marketing reporting is a dependable hedge against some safety risks, faster initial approval of new treatments and interventions may become possible.

Automated image interpretation has also been a promising subject of study for decades. Progress on interpreting large archives of weakly-labeled images, such as large photo archives scraped from the web, has been explosive. At first blush, it is surprising that there has not been a similar revolution in interpretation of medical images. Most medical imaging modalities (CT, MR, ultrasound) are inherently digital, the images are all archived, and there are large, established companies with internal R&D (e.g. Siemens, Philips, GE) devoted to imaging.

But several barriers have limited progress to date. Most hospital image archives have only gone digital over the past decade. More importantly, the problem in medicine is not to recognize what is in the image (is this a liver or a kidney?), but rather to make a fine-grained judgement about it (does the slightly darker smudge in the liver suggest a potentially cancerous tumor?). Strict regulations govern these high-stakes judgements. Even with state-of-the-art technologies, a radiologist will still likely have to look at the images, so the value proposition is not yet compelling. Also, healthcare regulations preclude easy federation of data across institutions. Thus, only very large organizations of integrated care, such as Kaiser Permanente, are able to attack these problems.

Still, automated/augmented image interpretation has started to gain momentum. The next fifteen years will probably not bring fully automated radiology, but initial forays into image "triage" or second level checking will likely improve the speed and cost-effectiveness of medical imaging. When coupled with electronic patient record systems, large-scale machine learning techniques could be applied to medical image data. For example, multiple major healthcare systems have archives of millions of patient scans, each of which has an associated radiological report, and most have an associated patient record. Already, papers are appearing in the literature showing that deep neural networks can be trained to produce basic radiological findings, with high reliability, by training from this data.[64]

> A small group of companies control the EHR market, and user interfaces are widely considered substandard, including annoying pop-ups that physicians routinely dismiss.

---

63   Sharecare, accessed August 1, 2016, *https://www.sharecare.com*.
64   Hoo-Chang Shin, Holger R. Roth, Mingchen Gao, Le Lu, Ziyue Xu, Isabella Nogues, Jianhua Yao, Daniel Mollura, and Ronald M. Summers, "Deep Convolutional Neural Networks for Computer-aided Detection: CNN Architectures, Dataset Characteristics and Transfer Learning," *IEEE Transactions on Medical Imaging* 35, no. 5 (2016): 1285–1298.

> The problem in medicine is not to recognize what is in the image—is this a liver or a kidney?—but rather to make a fine-grained judgement about it. Strict regulations govern these high-stakes judgements.

### Healthcare robotics

Fifteen years ago, healthcare robotics was largely science fiction. One company called Robodoc,[65] a spin-out from IBM, developed robotic systems for orthopedic surgeries, such as hip and knee replacements. The technology worked, but the company struggled commercially, and was ultimately shut down and acquired for its technology.[66] More recently, though, the research and practical use of surgical robotics has exploded.

In 2000 Intuitive Surgical[67] introduced the da Vinci system, a novel technology initially marketed to support minimally invasive heart bypass surgery, and then gained substantial market traction for treatment of prostate cancer and merged with its only major competition, Computer Motion, in 2003. The da Vinci, now in its fourth generation, provides 3D visualization (as opposed to 2D monocular laparoscopy) and wristed instruments in an ergonomic platform. It is considered the standard of care in multiple laparoscopic procedures, and used in nearly three quarters of a million procedures a year,[68] providing not only a physical platform, but also a new data platform for studying the process of surgery.

The da Vinci anticipates a day when much greater insight into how medical professionals carry out the process of providing interventional medical care will be possible. The presence of the da Vinci in day-to-day operation has also opened the doors to new types of innovation—from new instrumentation to image fusion to novel biomarkers—creating its own innovation ecosystem. The success of the platform has inspired potential competitors in robotic surgery, most notably the Alphabet spin-off Verb, in collaboration with J&J/Ethicon.[69] There are likely to be many more, each exploring a unique niche or space and building out an ecosystem of sensing, data analytics, augmentation, and automation.

Intelligent automation in hospital operations has been less successful. The story is not unlike surgical robotics. Twenty years ago, one company, HelpMate, created a robot for hospital deliveries,[70] such as meals and medical records, but ultimately went bankrupt. More recently, Aethon[71] introduced TUG Robots for basic deliveries, but few hospitals have invested in this technology to date. However, robotics in other service industries such as hotels and warehouses, including Amazon Robotics (formerly Kiva), are demonstrating that these technologies are practical and cost effective in at least some large-scale settings, and may ultimately spur additional innovation in health care.

Looking ahead, many tasks that appear in healthcare will be amenable to augmentation, but will not be fully automated. For example, robots may be able to deliver goods to the right room in a hospital, but then require a person to pick them up and place them in their final location. Walking a patient down the corridor may be relatively simple once a patient is standing in a walker (though will certainly not be trivial for patients recovering from surgery and/or elderly patients, especially in corridors crowded with equipment and other people). Driving a needle to place a suture is relatively straightforward once the needle is correctly placed.[72] This implies that many future systems will involve intimate interaction between people and machines and require technologies that facilitate collaboration between them.

---

65   ROBODOC, accessed August 1, 2016, http://www.robodoc.com/professionals.html.
66   THINK Surgical, accessed August 1, 2016, http://thinksurgical.com/history.
67   Intuitive Surgical, accessed August 1, 2016, http://www.intuitivesurgical.com.
68   Trefis Team, "Intuitive Surgical Maintains Its Growth Momentum With Strong Growth In Procedure Volumes," *Forbes*, January 22, 2016, accessed August 1, 2016, http://www.forbes.com/sites/greatspeculations/2016/01/22/intuitive-surgical-maintains-its-growth-momentum-with-strong-growth-in-procedure-volumes/#22ae6b0939a1.
69   Evan Ackerman, "Google and Johnson & Johnson Conjugate to Create Verb Surgical, Promise Fancy Medical Robots," *IEEE Spectrum*, December 17, 2015, accessed August 1, 2016, http://spectrum.ieee.org/automaton/robotics/medical-robots/google-verily-johnson-johnson-verb-surgical-medical-robots.
70   John M. Evans and Bala Krishnamurthy, "HelpMate®, the trackless robotic courier: A perspective on the development of a commercial autonomous mobile robot," *Lecture Notes in Control and Information Sciences 236*, June 18, 2005 (Springer-Verlag London Limited, 1998), 182–210, accessed August 1, 2016, http://link.springer.com/chapter/10.1007%2FBFb0030806.
71   Aethon, accessed August 1, 2016, http://www.aethon.com.

The growth of automation will enable new insights into healthcare processes. Historically, robotics has not been a strongly data-driven or data-oriented science. This is changing as (semi)automation infiltrates healthcare. As the new surgical, delivery, and patient care platforms come online, the beginnings of quantification and predictive analytics are being built on top of data coming from these platforms.[73] This data will be used to assess quality of performance, identify deficiencies, errors, or potential optimizations, and will be used as feedback to improve performance. In short, these platforms will facilitate making the connection between what is done, and the outcome achieved, making true "closed-loop" medicine a real possibility.

### Mobile health

To date, evidence-driven analytics on healthcare have relied on traditional healthcare data—mainly the electronic medical records discussed above. In the clinical setting, there are hopeful trends towards bringing new data to bear. For example, Tele-Language enables a human clinician to conduct language therapy sessions with multiple patients simultaneously with the aid of an AI agent trained by the clinician. And Lifegraph, which extracts behavioral patterns and creates alerts from data passively collected from a patient's smartphone, has been adopted by psychiatrists in Israel to detect early signs of distressful behavior in patients.

Looking ahead, driven by the mobile computing revolution, the astonishing growth of "biometrics in the wild"—and the explosion of platforms and applications that use them—is a hopeful and unanticipated trend. Thousands of mobile apps now offer information, introduce behavior modification, or identify groups of "people like me." This, combined with the emerging trend of more specialized motion tracking devices, such as Fitbit, and the emerging (inter)connectedness between the home environment and health-monitoring devices, has created a vibrant new sector of innovation.

By combining social and healthcare data, some healthcare apps can perform data mining, learning, and prediction from captured data, though their predictions are relatively rudimentary. The convergence of data and functionality across applications will likely spur new and even obvious products, such as an exercise app that not only proposes a schedule for exercise but also suggests the best time to do it, and provides coaching to stick to that schedule.

> **Specialized motion tracking devices… and the emerging (inter)connectedness between the home environment and health-monitoring devices have created a vibrant new sector of innovation.**

---

72   Azad Shademan, Ryan S. Decker, Justin D. Opfermann, Simon Leonard, Axel Krieger, and Peter CW Kim, "Supervised Autonomous Robotic Soft Tissue Surgery," *Science Translational Medicine* 8, no. 337 (2016): 337ra64–337ra64.
73   Carolyn Chen, Lee White, Timothy Kowalewski, Rajesh Aggarwal, Chris Lintott, Bryan Comstock, Katie Kuksenok, Cecilia Aragon, Daniel Holst, and Thomas Lendvay, "Crowd-Sourced Assessment of Technical Skills: a novel method to evaluate surgical performance." *Journal of Surgical Research* 187, no. 1 (2014): 65–71.

> Better hearing aids and visual assistive devices will mitigate the effects of hearing and vision loss, improving safety and social connection. Personalized rehabilitation and in-home therapy will reduce the need for hospital stays.

### Elder care

Over the next fifteen years the number of elderly in the United States will grow by over 50%.[74] The National Bureau of Labor Statistics projects that home health aides will grow 38% over the next ten years. Despite the broad opportunities in this domain—basic social support, interaction and communication devices, home health monitoring, a variety of simple in-home physical aids such as walkers, and light meal preparation—little has happened over the past fifteen years. But the coming generational shift will accompany a change in technology acceptance among the elderly. Currently, someone who is seventy was born in 1946 and may have first experienced some form of personalized IT in middle age or later, while a fifty-year-old today is far more technology-friendly and savvy. As a result, there will be a growing interest and market for already available and maturing technologies to support physical, emotional, social, and mental health. Here are a few likely examples by category:

**Life quality and independence**
- Automated transportation will support continued independence and expanded social horizons.
- Sharing of information will help families remain engaged with one another at a distance, and predictive analytics may be used to "nudge" family groups toward positive behaviors, such as reminders to "call home."
- Smart devices in the home will help with daily living activities when needed, such as cooking and, if robot manipulation capabilities improve sufficiently, dressing and toileting.

**Health and wellness**
- Mobile applications that monitor movement and activities, coupled with social platforms, will be able to make recommendations to maintain mental and physical health.
- In-home health monitoring and health information access will be able to detect changes in mood or behavior and alert caregivers.
- Personalized health management will help mitigate the complexities associated with multiple co-morbid conditions and/or treatment interactions.

**Treatments and devices**
- Better hearing aids and visual assistive devices will mitigate the effects of hearing and vision loss, improving safety and social connection.
- Personalized rehabilitation and in-home therapy will reduce the need for hospital or care facility stays.
- Physical assistive devices (intelligent walkers, wheel chairs, and exoskeletons) will extend the range of activities of an infirm individual.

The Study Panel expects an explosion of low-cost sensing technologies that can provide substantial capabilities to the elderly in their homes. In principle, social agents with a physical presence and simple physical capabilities (e.g. a mobile robot with basic communication capabilities) could provide a platform for new innovations. However, doing so will require integration across multiple areas of AI—Natural Language Processing, reasoning, learning, perception, and robotics—to create a system that is useful and usable by the elderly.

These innovations will also introduce questions regarding privacy within various circles, including friends, family, and care-givers, and create new challenges to accommodate an evermore active and engaged population far past retirement.

---

74  Jennifer M. Ortman, Victoria A. Velkoff, and Howard Hogan, "An Aging Nation: The Older Population in the United States: Population Estimates and Projections," *Current Population Reports*, U.S Census Bureau (May 2014), accessed August 1, 2016, https://www.census.gov/prod/2014pubs/p25-1140.pdf.

## EDUCATION

The past fifteen years have seen considerable AI advances in education. Applications are in wide use by educators and learners today, with some variation between K-12 and university settings. Though quality education will always require active engagement by human teachers, AI promises to enhance education at all levels, especially by providing personalization at scale. Similar to healthcare, resolving how to best integrate human interaction and face-to-face learning with promising AI technologies remains a key challenge.

Robots have long been popular educational devices, starting with the early Lego Mindstorms kits developed with the MIT Media Lab in the 1980s. Intelligent Tutoring Systems (ITS) for science, math, language, and other disciplines match students with interactive machine tutors. Natural Language Processing, especially when combined with machine learning and crowdsourcing, has boosted online learning and enabled teachers to multiply the size of their classrooms while simultaneously addressing individual students' learning needs and styles. The data sets from large online learning systems have fueled rapid growth in learning analytics.

Still, schools and universities have been slow in adopting AI technologies primarily due to lack of funds and lack of solid evidence that they help students achieve learning objectives. Over the next fifteen years in a typical North American city, the use of intelligent tutors and other AI technologies to assist teachers in the classroom and in the home is likely to expand significantly, as will learning based on virtual reality applications. But computer-based learning systems are not likely to fully replace human teaching in schools.

### Teaching robots

Today, more sophisticated and versatile kits for use in K-12 schools are available from a number of companies that create robots with new sensing technologies programmable in a variety of languages. Ozobot is a robot that teaches children to code and reason deductively while configuring it to dance or play based on color-coded patterns.[75] Cubelets help teach children logical thinking through assembling robot blocks to think, act, or sense, depending upon the function of the different blocks.[76] Wonder Workshop's Dash and Dot span a range of programming capabilities. Children eight years old and older can create simple actions using a visual programming language, Blockly, or build iOS and Android applications using C or Java.[77] PLEO rb is a robot pet that helps children learn biology by teaching the robot to react to different aspects of the environment.[78] However, while fun and engaging for some, in order for such kits to become widespread, there will need to be compelling evidence that they improve students' academic performance.

### Intelligent Tutoring Systems (ITS) and online learning

ITS have been developed from research laboratory projects such as Why-2 Atlas, which supported human-machine dialogue to solve physics problems early in the era.[79] The rapid migration of ITS from laboratory experimental stages to real use is

> Though quality education will always require active engagement by human teachers, AI promises to enhance education at all levels, especially by providing personalization at scale.

---

75  Ozobot, accessed August 1, 2016, http://ozobot.com/.
76  "Cubelets," Modular Robotics, accessed August 1, 2016, http://www.modrobotics.com/cubelets.
77  "Meet Dash," Wonder Workshop, accessed August 1, 2016, https://www.makewonder.com/dash.
78  "Pleo rb," Innvo Labs, accessed August 1, 2016, http://www.pleoworld.com/pleo_rb/eng/lifeform.php.
79  Kurt VanLehn, Pamela W. Jordan, Carolyn P. Rosé, Dumisizwe Bhembe, Michael Böttner, Andy Gaydos, Maxim Makatchev, Umarani Pappuswamy, Michael Ringenberg, Antonio Roque, Stephanie Siler, and Ramesh Srivastava, "The Architecture of Why2-Atlas: A Coach for Qualitative

> It can be argued that AI is the secret sauce that has enabled instructors, particularly in higher education, to multiply the size of their classrooms by a few orders of magnitude—class sizes of a few tens of thousands are not uncommon.

surprising and welcome. Downloadable software and online systems such as Carnegie Speech or Duolingo provide foreign language training using Automatic Speech Recognition (ASR) and NLP techniques to recognize language errors and help users correct them.[80] Tutoring systems such as the Carnegie Cognitive Tutor[81] have been used in US high schools to help students learn mathematics. Other ITS have been developed for training in geography, circuits, medical diagnosis, computer literacy and programming, genetics, and chemistry. Cognitive tutors use software to mimic the role of a good human tutor by, for example, providing hints when a student gets stuck on a math problem. Based on the hint requested and the answer provided, the tutor offers context specific feedback.

Applications are growing in higher education. An ITS called SHERLOCK[82] is beginning to be used to teach Air Force technicians to diagnose electrical systems problems in aircraft. And the University of Southern California's Information Sciences Institute has developed more advanced avatar-based training modules to train military personnel being sent to international posts in appropriate behavior when dealing with people from different cultural backgrounds. New algorithms for personalized tutoring, such as Bayesian Knowledge Tracing, enable individualized mastery learning and problem sequencing.[83]

Most surprising has been the explosion of the Massive Open Online Courses (MOOCs) and other models of online education at all levels—including the use of tools like Wikipedia and Khan Academy as well as sophisticated learning management systems that build in synchronous as well as asynchronous education and adaptive learning tools. Since the late 1990s, companies such as the Educational Testing Service and Pearson have been developing automatic NLP assessment tools to co-grade essays in standardized testing.[84] Many of the MOOCs which have become so popular, including those created by EdX, Coursera, and Udacity, are making use of NLP, machine learning, and crowdsourcing techniques for grading short-answer and essay questions as well as programming assignments.[85] Online education systems that support graduate-level professional education and lifelong learning are also expanding rapidly. These systems have great promise because the need for face-to-face interaction is less important for working professionals and career changers. While not the leaders in AI-supported systems and applications, they will become early adopters as the technologies are tested and validated.

It can be argued that AI is the secret sauce that has enabled instructors, particularly in higher education, to multiply the size of their classrooms by a few orders of magnitude—class sizes of a few tens of thousands are not uncommon. In order to continually test large classes of students, automated generation of the questions is

---

Physics Essay Writing," *Intelligent Tutoring Systems: Proceedings of the 6th International Conference*, (Springer Berlin Heidelberg, 2002), 158–167.
80   VanLehn et al, "The Architecture of Why2-Atlas."
81   "Resources and Support," Carnegie Learning, accessed August 1, 2016, https://www.carnegielearning.com/resources-support/.
82   Alan Lesgold, Suzanne Lajoie, Marilyn Bunzo, and Gary Eggan, "SHERLOCK: A Coached Practice Environment for an Electronics Troubleshooting Job," in J. H. Larkin and R. W. Chabay, eds., *Computer-Assisted Instruction and Intelligent Tutoring Systems: Shared Goals and Complementary Approaches* (Hillsdale, New Jersey: Lawrence Erlbaum Associates, 1988).
83   Michael V. Yudelson, Kenneth R. Koedinger, and Geoffrey J. Gordon, (2013). " Individualized Bayesian Knowledge Tracing Models," *Artificial Intelligence in Education*, (Springer Berlin Heidelberg, 2013), 171–180.
84   Jill Burstein, Karen Kukich, Susanne Wolff, Chi Lu, Martin Chodorow, Lisa Braden-Harder, and Mary Dee Harris, "Automated Scoring Using a Hybrid Feature Identification Technique" in *Proceedings of the Annual Meeting of the Association of Computational Linguistics*, Montreal, Canada, August 1998, accessed August 1, 2016, https://www.ets.org/Media/Research/pdf/erater_acl98.pdf.
85   EdX, https://www.edx.org/, Coursera, https://www.coursera.org/, Udacity, https://www.udacity.com/, all accessed August 1, 2016.

also possible, such as those designed to assess vocabulary,[86] wh (who/what/when/where/why) questions,[87] and multiple choice questions,[88] using electronic resources such as WordNet, Wikipedia, and online ontologies. With the explosion of online courses, these techniques are sure to be eagerly adopted for use in online education. Although the long term impact of these systems will have on the educational system remains unclear, the AI community has learned a great deal in a very short time.

### Learning analytics

Data sets being collected from massive scale online learning systems, ranging from MOOCs to Khan Academy, as well as smaller scale online programs, have fueled the rapid growth of the field of learning analytics. Online courses are not only good for widespread delivery, but are natural vehicles for data collection and experimental instrumentation that will contribute to scientific findings and improving the quality of learning at scale. Organizations such as the Society for Learning Analytics Research (SOLAR), and the rise of conferences including the Learning Analytics and Knowledge Conference[89] and the Learning at Scale Conference (L@S)[90] reflect this trend. This community applies deep learning, natural language processing, and other AI techniques to analysis of student engagement, behavior, and outcomes.

Current projects seek to model common student misconceptions, predict which students are at risk of failure, and provide real-time student feedback that is tightly integrated with learning outcomes. Recent work has also been devoted to understanding the cognitive processes involved in comprehension, writing, knowledge acquisition, and memory, and to applying that understanding to educational practice by developing and testing educational technologies.

### Challenges and opportunities

One might have expected more and more sophisticated use of AI technologies in schools, colleges, and universities by now. Much of its absence can be explained by the lack of financial resources of these institutions as well as the lack of data establishing the technologies' effectiveness. These problems are being addressed, albeit slowly, by private foundations and by numerous programs to train primarily secondary school teachers in summer programs. As in other areas of AI, excessive hype and promises about the capabilities of MOOCs have meant that expectations frequently exceed the reality. The experiences of certain institutions, such as San Jose State University's experiment with Udacity,[91] have led to more sober assessment of the potential of the new educational technologies.

**The current absence of sophisticated use of AI technologies in schools, colleges, and universities may be explained by the lack of financial resources as well as the lack of data establishing the technologies' effectiveness.**

---

86   Jonathan C. Brown, Gwen A. Frishkoff, and Maxine Eskenazi, "Automatic Question Generation for Vocabulary Assessment," *Proceedings of Human Language Technology Conference and Conference on Empirical Methods in Natural Language Processing (HLT/EMNLP)*, Vancouver, October 2005, (Association for Computational Linguistics, 2005), 819–826.
87   Michael Heilman, "Automatic Factual Question Generation from Text," PhD thesis CMU-LTI-11-004, (Carnegie Mellon University, 2011), accessed August 1, 2016, http://www.cs.cmu.edu/~ark/mheilman/questions/papers/heilman-question-generation-dissertation.pdf.
88   Tahani Alsubait, Bijan Parsia, and Uli Sattler, "Generating Multiple Choice Questions from Ontologies: How Far Can We Go?," in eds. P. Lambrix, E. Hyvönen. E. Blomqvist, V. Presutti, G. Qi, U. Sattler, Y. Ding, and C. Ghidini, *Knowledge Engineering and Knowledge Management: EKAW 2014 Satellite Events, VISUAL, EKM1, and ARCOE-Logic* Linköping, Sweden, November 24–28, 2014 Revised Selected Papers, (Switzerland: Springer International Publishing, 2015), 66–79.
89   The 6th International Learning Analytics & Knowledge Conference, accessed August 1, 2016, http://lak16.solaresearch.org/.
90   Third Annual ACM Conference on Learning at Scale, http://learningatscale.acm.org/las2016/.
91   Ry Rivard, "Udacity Project on 'Pause'," *Inside Higher Ed*, July 18, 2013, accessed August 1, 2016, https://www.insidehighered.com/news/2013/07/18/citing-disappointing-student-outcomes-san-jose-state-pauses-work-udacity.

> While formal education will not disappear, the Study Panel believes that MOOCs and other forms of online education will become part of learning at all levels, from K-12 through university, in a blended classroom experience.

In the next fifteen years, it is likely that human teachers will be assisted by AI technologies with better human interaction, both in the classroom and in the home. The Study Panel expects that more general and more sophisticated virtual reality scenarios in which students can immerse themselves in subjects from all disciplines will be developed. Some steps in this direction are being taken now by increasing collaborations between AI researchers and researchers in the humanities and social sciences, exemplified by Stanford's Galileo Correspondence Project[92] and Columbia's Making and Knowing Project.[93] These interdisciplinary efforts create interactive experiences with historical documents and the use of Virtual Reality (VR) to explore interactive archeological sites.[94] VR techniques are already being used in the natural sciences such as biology, anatomy, geology and astronomy to allow students to interact with environments and objects that are difficult to engage with in the real world. The recreation of past worlds and fictional worlds will become just as popular for studies of arts and other sciences.

AI techniques will increasingly blur the line between formal, classroom education and self-paced, individual learning. Adaptive learning systems, for example, are going to become a core part of the teaching process in higher education because of the pressures to contain cost while serving a larger number of students and moving students through school more quickly. While formal education will not disappear, the Study Panel believes that MOOCs and other forms of online education will become part of learning at all levels, from K-12 through university, in a blended classroom experience. This development will facilitate more customizable approaches to learning, in which students can learn at their own pace using educational techniques that work best for them. Online education systems will learn as the students learn, supporting rapid advances in our understanding of the learning process. Learning analytics, in turn, will accelerate the development of tools for personalized education.

The current transition from hard copy books to digital and audio media and texts is likely to become prevalent in education as well. Digital reading devices will also become much 'smarter', providing students with easy access to additional information about subject matter as they study. Machine Translation (MT) technology will also make it easier to translate educational material into different languages with a fair degree of accuracy, just as it currently translates technical manuals. Textbook translation services that currently depend only upon human translators will increasingly incorporate automatic methods to improve the speed and affordability of their services for school systems.

Online learning systems will also expand the opportunity for adults and working professionals to enhance their knowledge and skills (or to retool and learn a new field) in a world where these fields are evolving rapidly. This will include the expansion of fully online professional degrees as well as professional certifications based on online coursework.

### Broader societal consequences

In countries where education is difficult for the broad population to obtain, online resources may have a positive effect if the population has the tools to access them. The development of online educational resources should make it easier for foundations that support international educational programs to provide quality education by providing tools and relatively simple amounts of training in their use. For example, large numbers of educational apps, many of them free, are being developed for the iPad. On the negative side, there is already a major trend among students to restrict their social contacts to electronic ones and to spend large amounts of time without social contact, interacting with online programs. If education also occurs more and more online, what effect will the lack of regular, face-to-face contact with peers have on students' social development? Certain technologies have even been shown to create neurological side effects.[95] On the other hand, autistic children have benefited from interactions with AI systems already.[96]

---

92  Stanford University: Galileo Correspondence Project, accessed August 1, 2016, http://galileo.stanford.edu.
93  The Making and Knowing Project: Reconstructing the 16th Century Workshop of BNF MS. FR. 640 at Columbia University, accessed August 1, 2016, http://www.makingandknowing.org.
94  Paul James, "3D Mapped HTC Vive Demo Brings Archaeology to Life," *Road to VR*, August 31, 2015, accessed August 1, 2016, http://www.roadtovr.com/3d-mapped-htc-vive-demo-brings-archaeology-to-life/.

## LOW-RESOURCE COMMUNITIES

Many opportunities exist for AI to improve conditions for people in low-resource communities in a typical North American city—and, indeed, in some cases it already has. Understanding these direct AI contributions may also inform potential contributions in the poorest parts of the developing world. There has not been a significant focus on these populations in AI gatherings, and, traditionally, AI funders have underinvested in research lacking commercial application. With targeted incentives and funding priorities, AI technologies could help address the needs of low-resource communities. Budding efforts are promising. Counteracting fears that AI may contribute to joblessness and other societal problems, AI may provide mitigations and solutions, particularly if implemented in ways that build trust in them by the affected communities.

**Machine learning, data mining approaches**

Under the banner of "data science for social good," AI has been used to create predictive models to help government agencies more effectively use their limited budgets to address problems such as lead poisoning,[97] a major public health concern that has been in the news due to ongoing events in Flint, Michigan. Children may be tested for elevated lead levels, but that unfortunately means the problem is only detected after they have already been poisoned. Many efforts are underway to use predictive models to assist government agencies in prioritizing children at risk, including those who may not yet have been exposed.[98] Similarly, the Illinois Department of Human Services (IDHS) uses predictive models to identify pregnant women at risk for adverse birth outcomes in order to maximize the impact of prenatal care. The City of Cincinnati uses them to proactively identify and deploy inspectors to properties at risk of code violations.

**Scheduling, planning**

Task assignment scheduling and planning techniques have been applied by many different groups to distribute food before it spoils from those who may have excess, such as restaurants, to food banks, community centers and individuals.[99]

> With targeted incentives and funding priorities, AI technologies could help address the needs of low-resource communities. Budding efforts are promising.

---

95   Scientist have studied, for example, the way reliance on GPS may lead to changes in the hippocampus. Kim Tingley, "The Secrets of the Wave Pilots," *The New York Times*, March 17, 2016, accessed August 1, 2016, http://www.nytimes.com/2016/03/20/magazine/the-secrets-of-the-wave-pilots.html.
96   Judith Newman, "To Siri, With Love: How One Boy With Autism Became BFF With Apple's Siri," *The New York Times*, October 17, 2014, accessed August 1, 2016, http://www.nytimes.com/2014/10/19/fashion/how-apples-siri-became-one-autistic-boys-bff.html.
97   Eric Potash, Joe Brew, Alexander Loewi, Subhabrata Majumdar, Andrew Reece, Joe Walsh, Eric Rozier, Emile Jorgensen, Raed Mansour, and Rayid Ghani, "Predictive Modeling for Public Health: Preventing Childhood Lead Poisoning," *Proceedings of the 21th ACM SIGKDD International Conference on Knowledge Discovery and Data Mining* (New York: Association for Computing Machinery, 2015), 2039–2047.
98   Data Science for Social Good, University of Chicago, accessed August 1, 2016, http://dssg.uchicago.edu/.
99   Senay Solak, Christina Scherrer, and Ahmed Ghoniem, "The Stop-and-Drop Problem in Nonprofit Food Distribution Networks," *Annals of Operations Research 221*, no. 1 (October 2014):

> One of the more successful uses of AI analytics is in detecting white collar crime, such as credit card fraud. Cybersecurity (including spam) is a widely shared concern, and machine learning is making an impact.

**Reasoning with social networks and influence maximization**
Social networks can be harnessed to create earlier, less-costly interventions involving large populations. For example, AI might be able to assist in spreading health-related information. In Los Angeles, there are more than 5,000 homeless youth (ages thirteen-twenty-four). Individual interventions are difficult and expensive, and the youths' mistrust of authority dictates that key messages are best spread through peer leaders. AI programs might be able to leverage homeless youth social networks to strategically select peer leaders to spread health-related information, such as how to avoid spread of HIV. The dynamic, uncertain nature of these networks does pose challenges for AI research.[100] Care must also be taken to prevent AI systems from reproducing discriminatory behavior, such as machine learning that identifies people through illegal racial indicators, or through highly-correlated surrogate factors, such as zip codes. But if deployed with great care, greater reliance on AI may well result in a reduction in discrimination overall, since AI programs are inherently more easily audited than humans.

## PUBLIC SAFETY AND SECURITY

Cities already have begun to deploy AI technologies for public safety and security. By 2030, the typical North American city will rely heavily upon them. These include cameras for surveillance that can detect anomalies pointing to a possible crime, drones, and predictive policing applications. As with most issues, there are benefits and risks. Gaining public trust is crucial. While there are legitimate concerns that policing that incorporates AI may become overbearing or pervasive in some contexts, the opposite is also possible. AI may enable policing to become more targeted and used only when needed. And assuming careful deployment, AI may also help remove some of the bias inherent in human decision-making.

One of the more successful uses of AI analytics is in detecting white collar crime, such as credit card fraud.[101] Cybersecurity (including spam) is a widely shared concern, and machine learning is making an impact. AI tools may also prove useful in helping police manage crime scenes or search and rescue events by helping commanders prioritize tasks and allocate resources, though these tools are not yet ready for automating such activities. Improvements in machine learning in general, and transfer learning in particular—for speeding up learning in new scenarios based on similarities with past scenarios—may facilitate such systems.

The cameras deployed almost everywhere in the world today tend to be more useful for helping solve crimes than preventing them.[102] [103] This is due to the low quality of event identification from videos and the lack of manpower to look at massive video streams. As AI for this domain improves, it will better assist crime prevention and prosecution through greater accuracy of event classification and efficient automatic processing of video to detect anomalies—including, potentially,

---

407–426.
100   Jordan Pearson, "Artificial Intelligence Could Help Reduce HIV Among Homeless Youths," Teamcore, University of Southern California, February 4. 2015, accessed August 1, 2016, http://teamcore.usc.edu/news/motherboard_news_ai_could_help_reduce_HIV.pdf.
101   "RSA Adaptive Authentication," *RSA*, accessed August 1, 2016, https://www.rsa.com/en-us/products-services/fraud-prevention/adaptive-authentication.
102   Takeshi Arikuma and Yasunori Mochizuki, "Intelligent multimedia surveillance system for safer cities" *APSIPA Transactions on Signal and Information Processing 5* (2016): 1–8.
103   "Big Op-Ed: Shifting Opinions On Surveillance Cameras,", *Talk of the Nation*, NPR, April 22, 2013, accessed August 1, 2016, http://www.npr.org/2013/04/22/178436355/big-op-ed-shifting-opinions-on-surveillance-cameras.

evidence of police malpractice. These improvements could lead to even more widespread surveillance. Some cities have already added drones for surveillance purposes, and police use of drones to maintain security of ports, airports, coastal areas, waterways, industrial facilities is likely to increase, raising concerns about privacy, safety, and other issues.

The New York Police Department's CompStat was the first tool pointing toward predictive policing,[104] and many police departments now use it.[105] Machine learning significantly enhances the ability to predict where and when crimes are more likely to happen and who may commit them. As dramatized in the movie *Minority Report*, predictive policing tools raise the specter of innocent people being unjustifiably targeted. But well-deployed AI prediction tools have the potential to actually remove or reduce human bias, rather than reinforcing it, and research and resources should be directed toward ensuring this effect.

AI techniques can be used to develop intelligent simulations for training law-enforcement personnel to collaborate. While international criminal organizations and terrorists from different countries are colluding, police forces from different countries still face difficulty in joining forces to fight them. Training international groups of law enforcement personnel to work as teams is very challenging. The European Union, through the Horizon 2020 program, currently supports such attempts in projects such as LawTrain.[106] The next step will be to move from simulation to actual investigations by providing tools that support such collaborations.

Tools do exist for scanning Twitter and other feeds to look for certain types of events and how they may impact security. For example, AI can help in social network analysis to prevent those at risk from being radicalized by ISIS or other violent groups. Law enforcement agencies are increasingly interested in trying to detect plans for disruptive events from social media, and also to monitor activity at large gatherings of people to analyze security. There is significant work on crowd simulations to determine how crowds can be controlled. At the same time, legitimate concerns have been raised about the potential for law enforcement agencies to overreach and use such tools to violate people's privacy.

The US Transportation Security Administration (TSA), Coast Guard, and the many other security agencies that currently rely on AI will likely increase their reliance to enable significant efficiency and efficacy improvements.[107] AI techniques—vision, speech analysis, and gait analysis— can aid interviewers, interrogators, and security guards in detecting possible deception and criminal behavior. For example, the TSA currently has an ambitious project to redo airport security nationwide.[108] Called DARMS, the system is designed to improve efficiency and efficacy of airport security by relying on personal information to tailor security based on a person's risk categorization and the flights being taken. The future vision for this project is a tunnel that checks people's security while they walk through it. Once again, developers of this technology should be careful to avoid building in bias (e.g. about a person's risk level category) through use of datasets that reflect prior bias.[109]

> As dramatized in the movie *Minority Report*, predictive policing tools raise the specter of innocent people being unjustifiably targeted. But well-deployed AI prediction tools have the potential to actually remove or reduce human bias.

---104  Walter L. Perry, Brian McInnis, Carter C. Price, Susan Smith, and John S. Hollywood, "The Role of Crime Forecasting in Law Enforcement Operations," *Rand Corporation Report 233* (2013).
105  "CompStat," *Wikipedia*, last modified July 28, 2016, accessed August 1, 2016, https://en.wikipedia.org/wiki/CompStat.
106  LAW-TRAIN, accessed August 1, 2016, http://www.law-train.eu/.
107  Milind Tambe, *Security and Game Theory: Algorithms, Deployed Systems, Lessons Learned* (New York: Cambridge University Press, 2011).
108  Peter Neffenger, "TSA's 2017 Budget—A Commitment to Security (Part I)," Department of Homeland Security, March 1, 2016, accessed August 1, 2016, https://www.tsa.gov/news/testimony/2016/03/01/hearing-fy17-budget-request-transportation-security-administration.
109  Crawford, "AI's White Guy Problem."

## EMPLOYMENT AND WORKPLACE

While AI technologies are likely to have a profound future impact on employment and workplace trends in a typical North American city, it is difficult to accurately assess current impacts, positive or negative. In the past fifteen years, employment has shifted due to a major recession and increasing globalization, particularly with China's introduction to the world economy, as well as enormous changes in non-AI digital technology. Since the 1990s, the US has experienced continued growth in productivity and GDP, but median income has stagnated and the employment to population ratio has fallen.

There are clear examples of industries in which digital technologies have had profound impacts, good and bad, and other sectors in which automation will likely make major changes in the near future. Many of these changes have been driven strongly by "routine" digital technologies, including enterprise resource planning, networking, information processing, and search. Understanding these changes should provide insights into how AI will affect future labor demand, including the shift in skill demands. To date, digital technologies have been affecting workers more in the skilled middle, such as travel agents, rather than the very lowest-skilled or highest skilled work.[110] On the other hand, the spectrum of tasks that digital systems can do is evolving as AI systems improve, which is likely to gradually increase the scope of what is considered routine. AI is also creeping into high end of the spectrum, including professional services not historically performed by machines.

To be successful, AI innovations will need to overcome understandable human fears of being marginalized. AI will likely replace tasks rather than jobs in the near term, and will also create new kinds of jobs. But the new jobs that will emerge are harder to imagine in advance than the existing jobs that will likely be lost. Changes in employment usually happen gradually, often without a sharp transition, a trend likely to continue as AI slowly moves into the workplace. A spectrum of effects will emerge, ranging from small amounts of replacement or augmentation to complete replacement. For example, although most of a lawyer's job is not yet automated,[111] AI applied to legal information extraction and topic modeling has automated parts of first-year lawyers' jobs.[112] In the not too distant future, a diverse array of job-holders, from radiologists to truck drivers to gardeners, may be affected.

AI may also influence the size and location of the workforce. Many organizations and institutions are large because they perform functions that can be scaled only by adding human labor, either "horizontally" across geographical areas or "vertically" in management hierarchies. As AI takes over many functions, scalability no longer implies large organizations. Many have noted the small number of employees of some high profile internet companies, but not of others. There may be a natural scale of human enterprise, perhaps where the CEO can know everyone in the company. Through the creation of efficiently outsourced labor markets enabled by AI, enterprises may tend towards that natural size.

AI will also create jobs, especially in some sectors, by making certain tasks more important, and create new categories of employment by making new modes of interaction possible. Sophisticated information systems can be used to create new

> AI will likely replace tasks rather than jobs in the near term, and will also create new kinds of jobs. But the new jobs that will emerge are harder to imagine in advance than the existing jobs that will likely be lost.

---

110  Jeremy Ashkenas and Alicia Parlapiano, "How the Recession Reshaped the Economy, in 255 Charts," *The New York Times*, June 6, 2014, accessed August 1, 2016, http://www.nytimes.com/interactive/2014/06/05/upshot/how-the-recession-reshaped-the-economy-in-255-charts.html.

111  R Dana Remus and Frank S. Levy, "Can Robots Be Lawyers? Computers, Lawyers, and the Practice of Law," *Social Science Research Network*, last modified February 12, 2016, accessed August 1, 2016, http://papers.ssrn.com/sol3/papers.cfm?abstract_id=2701092.

112  John Markoff, "Armies of Expensive Lawyers, Replaced by Cheaper Software," *The New York Times*, March 4, 2011, accessed August 1, 2016, http://www.nytimes.com/2011/03/05/science/05legal.html.

markets, which often have the effect of lowering barriers to entry and increasing participation—from app stores to AirBnB to taskrabbit. A vibrant research community within AI studies further ways of creating new markets and making existing ones operate more efficiently.

While work has intrinsic value, most people work to be able to purchase goods and services they value. Because AI systems perform work that previously required human labor, they have the effect of lowering the cost of many goods and services, effectively making everyone richer. But as exemplified in current political debates, job loss is more salient to people—especially those directly affected—than diffuse economic gains, and AI unfortunately is often framed as a threat to jobs rather than a boon to living standards.

There is even fear in some quarters that advances in AI will be so rapid as to replace all human jobs—including those that are largely cognitive or involve judgment—within a single generation. This sudden scenario is highly unlikely, but AI will gradually invade almost all employment sectors, requiring a shift away from human labor that computers are able to take over.

The economic effects of AI on cognitive human jobs will be analogous to the effects of automation and robotics on humans in manufacturing jobs. Many middle-aged workers have lost well-paying factory jobs and the socio-economic status in family and society that traditionally went with such jobs. An even larger fraction of the total workforce may, in the long run, lose well-paying "cognitive" jobs. As labor becomes a less important factor in production as compared to owning intellectual capital, a majority of citizens may find the value of their labor insufficient to pay for a socially acceptable standard of living. These changes will require a political, rather than a purely economic, response concerning what kind of social safety nets should be in place to protect people from large, structural shifts in the economy. Absent mitigating policies, the beneficiaries of these shifts may be a small group at the upper stratum of the society.[113]

In the short run, education, re-training, and inventing new goods and services may mitigate these effects. Longer term, the current social safety net may need to evolve into better social services for everyone, such as healthcare and education, or a guaranteed basic income. Indeed, countries such as Switzerland and Finland have actively considered such measures. AI may be thought of as a radically different mechanism of wealth creation in which everyone should be entitled to a portion of the world's AI-produced treasure.[114] It is not too soon for social debate on how the economic fruits of AI-technologies should be shared. As children in traditional societies support their aging parents, perhaps our artificially intelligent "children" should support us, the "parents" of their intelligence.

> As labor becomes a less important factor in production as compared to owning intellectual capital, a majority of citizens may find the value of their labor insufficient to pay for a socially acceptable standard of living.

---

113  For example, Brynjolfsson and McAfee, *Second Machine Age*, have two chapters of devoted to this (Erik Brynjolfsson and Andrew McAfee, *The Second Machine Age: Work, Progress, and Prosperity in a Time of Brilliant Technologies*, (New York: W. W. Norton & Company, Inc., 2014)) and Brynjolfsson, McAfee, and Spence describe policy responses for the combination of globalization and digital technology (Erik Brynjolfsson, Andrew McAfee, and Michael Spence, *Foreign Affairs*, July/August 2014, accessed August 1, 2016, https://www.foreignaffairs.com/articles/united-states/2014-06-04/new-world-order).

114  GDP does not do a good job of measuring the value of many digital goods. When society can't manage what isn't measured, bad policy decisions result. One alternative is to look at consumer surplus, not just dollar flows. As AI is embodied in more goods, this issue becomes more salient. It may look like GDP goes down but people have better well-being through access to these digital goods. See Erik Brynjolfsson and Adam Saunders, "What the GDP Gets Wrong (Why Managers Should Care)," *Sloan Management Review*, vol. 51, no. 1 (October 1, 2009): 95–96.

## ENTERTAINMENT

With the explosive growth of the internet over the past fifteen years, few can imagine their daily lives without it. Powered by AI, the internet has established user-generated content as a viable source of information and entertainment. Social networks such as Facebook are now pervasive, and they function as personalized channels of social interaction and entertainment—sometimes to the detriment of interpersonal interaction. Apps such as WhatsApp and Snapchat enable smart-phone users to remain constantly "in touch" with peers and share sources of entertainment and information. In on-line communities such as Second Life and role-playing games such as World of Warcraft, people imagine an alternative existence in a virtual world.[115] Specialized devices, such as Amazon's Kindle have also redefined the essentials of long-cherished pastimes. Books can now be browsed and procured with a few swipes of the finger, stored by the thousands in a pocket-sized device, and read in much the same way as a handheld paperback.

Trusted platforms now exist for sharing and browsing blogs, videos, photos, and topical discussions, in addition to a variety of other user-generated information. To operate at the scale of the internet, these platforms must rely on techniques that are being actively developed in natural language processing, information retrieval, image processing, crowdsourcing, and machine learning. Algorithms such as collaborative filtering have been developed, for example, to recommend relevant movies, songs, or articles based on the user's demographic details and browsing history.[116]

Traditional sources of entertainment have also embraced AI to keep pace with the times. As exemplified in the book and movie *Moneyball*, professional sport is now subjected to intensive quantitative analysis.[117] Beyond aggregate performance statistics, on-field signals can be monitored using sophisticated sensors and cameras. Software has been created for composing music[118] and recognizing soundtracks.[119] Techniques from computer vision and NLP have been used in creating stage performances.[120] Even the lay user can exercise his or her creativity on platforms such as WordsEye, which automatically generates 3D scenes from natural language text.[121] AI has also come to the aid of historical research in the arts, and is used extensively in stylometry and, more recently, in the analysis of paintings.[122]

The enthusiasm with which humans have responded to AI-driven entertainment has been surprising and led to concerns that it reduces interpersonal interaction among human beings. Few predicted that people would spend hours on end interacting with a display. Children often appear to be genuinely happier playing at home on their devices rather than outside with their friends. AI will increasingly enable entertainment that is more interactive, personalized, and engaging. Research should be directed toward understanding how to leverage these attributes for individuals' and society's benefit.

> AI will increasingly enable entertainment that is more interactive, personalized, and engaging. Research should be directed toward understanding how to leverage these attributes for individuals' and society's benefit.

---

115  Second Life, accessed August 1, 2016, http://secondlife.com; "World of Warcraft," Blizzard Entertainment, Inc, accessed August 1, 2016, http://us.battle.net/wow/en/.
116  John S. Breese, David Heckerman, and Carl Kadie, "Empirical Analysis of Predictive Algorithms for Collaborative Filtering," *Proceedings of the 14th Conference on Uncertainty in Artificial Intelligence* (July 1998), accessed August 1, 2016, http://arxiv.org/pdf/1301.7363.pdf, 43–52.
117  Michael Lewis, *Moneyball: The Art of Winning an Unfair Game* (New York: W. W. Norton & Company, Inc., 2003): http://www.imdb.com/title/tt1210166/).
118  MuseScore, accessed August 1, 2016, https://musescore.org/.
119  Shazam, accessed August 1, 2016, http://www.shazam.com/.
120  Annie Dorsen, accessed August 1, 2016, http://www.anniedorsen.com/.
121  WordsEye, accessed August 1, 2016, https://www.wordseye.com/.
122  "Stylometry," *Wikipedia*, last modified August 4, 2016, accessed August 1, 2016, https://en.wikipedia.org/wiki/Stylometry; http://arxiv.org/pdf/1408.3218v1.pdf.

## Imagining the Future

The success of any form of entertainment is ultimately determined by the individuals and social groups that are its subjects. The modes of entertainment that people find appealing are diverse and change over time. It is therefore hard to predict the forms entertainment will take in the next fifteen years precisely. Nevertheless, current trends suggest at least a few features that the future entertainment landscape is likely to contain.

To date, the information revolution has mostly unfolded in software. However, with the growing availability of cheaper sensors and devices, greater innovation in the hardware used in entertainment systems is expected. Virtual reality and haptics could enter our living rooms—personalized companion robots are already being developed.[123] With the accompanying improvements in Automatic Speech Recognition, the Study Panel expects that interaction with robots and other entertainment systems will become dialogue-based, perhaps constrained at the start, but progressively more human-like. Equally, the interacting systems are predicted to develop new characteristics such as emotion, empathy, and adaptation to environmental rhythms such as time of day.[124]

Today, an amateur with a video camera and readily-available software tools can make a relatively good movie. In the future, more sophisticated tools and apps will become available to make it even easier to produce high-quality content, for example, to compose music or to choreograph dance using an avatar. The creation and dissemination of entertainment will benefit from the progress of technologies such as ASR, dubbing, and Machine Translation, which will enable content to be customized to different audiences inexpensively. This democratization and proliferation of AI-created media makes it difficult to predict how humans' taste for entertainment, which are already fluid, will evolve.

With content increasingly delivered digitally, and large amounts of data being logged about consumers' preferences and usage characteristics, media powerhouses will be able to micro-analyze and micro-serve content to increasingly specialized segments of the population—down to the individual.[125] Conceivably the stage is set for the emergence of media conglomerates acting as "Big Brothers" who are able to control the ideas and online experiences to which specific individuals are exposed. It remains to be seen whether broader society will develop measures to prevent their emergence. This topic, along with others pertaining to AI-related policy, is treated in more detail in the next section.

> **More sophisticated tools and apps will become available to make it even easier to produce high-quality content, for example, to compose music or to choreograph dance using an avatar.**

---

123   Emoters, accessed August 1, 2016, http://emoterbots.com/.
124   "Siri," Apple, Inc., accessed August 1, 2016, http://www.apple.com/in/ios/siri/.
125   Ryan Calo, "Digital Market Manipulation," *George Washington Law Review* 82, no. 4 (2014): 995–1051.

# SECTION III: PROSPECTS AND RECOMMENDATIONS FOR AI PUBLIC POLICY

*The goal of AI applications must be to create value for society. Our policy recommendations flow from this goal, and, while this report is focused on a typical North American city in 2030, the recommendations are broadly applicable to other places over time. Strategies that enhance our ability to interpret AI systems and participate in their use may help build trust and prevent drastic failures. Care must be taken to augment and enhance human capabilities and interaction, and to avoid discrimination against segments of society. Research to encourage this direction and inform public policy debates should be emphasized. Given the current sector-specific regulation of US industries, new or retooled laws and policies will be needed to address the widespread impacts AI is likely to bring. Rather than "more" or "stricter" regulation, policies should be designed to encourage helpful innovation, generate and transfer expertise, and foster broad corporate and civic responsibility for addressing critical societal issues raised by these technologies. In the long term, AI will enable new wealth creation that will require social debate on how the economic fruits of AI technologies should be shared.*

## AI POLICY, NOW AND IN THE FUTURE

Throughout history, humans have both shaped and adapted to new technologies. This report anticipates that advances in AI technologies will be developed and fielded gradually—not in sudden, unexpected jumps in the techniques themselves—and will build on what exists today, making this adaptation easier. On the other hand, small improvements to techniques, computing power, or availability of data can occasionally lead to novel, game-changing applications. The measure of success for AI applications is the value they create for human lives. Going forward, the ease with which people use and adapt to AI applications will likewise largely determine their success.

Conversely, since AI applications are susceptible to errors and failures, a mark of their success will be how users perceive and tolerate their shortcomings. As AI becomes increasingly embedded in daily lives and used for more critical tasks, system mistakes may lead to backlash from users and negatively affect their trust. Though accidents in a self-driving car may be less probable than those driven by humans, for example, they will attract more attention. Design strategies that enhance the ability of humans to understand AI systems and decisions (such as explicitly explaining those decisions), and to participate in their use, may help build trust and prevent drastic failures. Likewise, developers should help manage people's expectations, which will affect their happiness and satisfaction with AI applications. Frustration in carrying out functions promised by a system diminishes people's trust and reduces their willingness to use the system in the future.

Another important consideration is how AI systems that take over certain tasks will affect people's affordances and capabilities. As machines deliver super-human performances on some tasks, people's ability to perform them may wither. Already, introducing calculators to classrooms has reduced children's ability to do basic arithmetic operations. Still, humans and AI systems have complementary abilities. People are likely to focus on tasks that machines cannot do as well, including complex reasoning and creative expression.

Already, children are increasingly exposed to AI applications, such as interacting with personal assistants on cell phones or with virtual agents in theme parks. Having early exposure will improve children's interactions with AI applications, which will become a natural part of their daily lives. As a result, gaps will appear in how younger and older generations perceive AI's influences on society.

> The measure of success for AI applications is the value they create for human lives. Going forward, the ease with which people use and adapt to AI applications will likewise largely determine their success.

Likewise, AI could widen existing inequalities of opportunity if access to AI technologies—along with the high-powered computation and large-scale data that fuel many of them—is unfairly distributed across society. These technologies will improve the abilities and efficiency of people who have access to them. A person with access to accurate Machine Translation technology will be better able to use learning resources available in different languages. Similarly, if speech translation technology is only available in English, people who do not speak English will be at a disadvantage.

Further, AI applications and the data they rely upon may reflect the biases of their designers and users, who specify the data sources. This threatens to deepen existing social biases, and concentrate AI's benefits unequally among different subgroups of society. For example, some speech recognition technologies do not work well for women and people with accents. As AI is increasingly used in critical applications, these biases may surface issues of fairness to diverse groups in society. On the other hand, compared to the well-documented biases in human decision-making, AI-based decision-making tools have the potential to significantly reduce the bias in critical decisions such as who is lent money or sent to jail.

Privacy concerns about AI-enabled surveillance are also widespread, particularly in cities with pervasive instrumentation. Sousveillance, the recording of an activity by a participant, usually with portable personal devices, has increased as well. Since views about bias and privacy are based on personal and societal ethical and value judgments, the debates over how to address these concerns will likely grow and resist quick resolution. Similarly, since AI is generating significant wealth, debates will grow regarding how the economic fruits of AI technologies should be shared—especially as AI expertise and the underlying data sets that fuel applications are concentrated in a small number of large corporations.

To help address these concerns about the individual and societal implications of rapidly evolving AI technologies, the Study Panel offers three general policy recommendations:

1. **Define a path toward accruing technical expertise in AI at all levels of government. Effective governance requires more experts who understand and can analyze the interactions between AI technologies, programmatic objectives, and overall societal values.**

Absent sufficient technical expertise to assess safety or other metrics, national or local officials may refuse to permit a potentially promising application. Or insufficiently trained officials may simply take the word of industry technologists and green light a sensitive application that has not been adequately vetted. Without an understanding of how AI systems interact with human behavior and societal values, officials will be poorly positioned to evaluate the impact of AI on programmatic objectives.

2. **Remove the perceived and actual impediments to research on the fairness, security, privacy, and social impacts of AI systems.**

Some interpretations of federal laws such as the Computer Fraud and Abuse Act and the anti-circumvention provision of the Digital Millennium Copyright Act are ambiguous regarding whether and how proprietary AI systems may be reverse engineered and evaluated by academics, journalists, and other researchers. Such research is critical if AI systems with physical and other material consequences are to be properly vetted and held accountable.

3. **Increase public and private funding for interdisciplinary studies of the societal impacts of AI.**

As a society, we are underinvesting resources in research on the societal implications of AI technologies. Private and public dollars should be directed toward interdisciplinary

> AI could widen existing inequalities of opportunity if access to AI technologies—along with the high-powered computation and large-scale data that fuel many of them—is unfairly distributed across society.

> As a society, we are underinvesting resources in research on the societal implications of AI technologies. Private and public dollars should be directed toward interdisciplinary teams capable of analyzing AI from multiple angles.

teams capable of analyzing AI from multiple angles. Research questions range from basic research into intelligence to methods to assess and affect the safety, privacy, fairness, and other impacts of AI.

Questions include: Who is responsible when a self-driven car crashes or an intelligent medical device fails? How can AI applications be prevented from unlawful discrimination? Who should reap the gains of efficiencies enabled by AI technologies and what protections should be afforded to people whose skills are rendered obsolete? As AI becomes integrated more broadly and deeply into industrial and consumer products, it enters areas in which established regulatory regimes will need to be adapted to AI innovations or in some cases fundamentally reconfigured according to broadly accepted goals and principles.

The approach in the United States to date has been sector-specific, with oversight by a variety of agencies. The use of AI in devices that deliver medical diagnostics and treatments is subject to aggressive regulation by the Food and Drug Administration (FDA), both in defining what the product is and specifying the methods by which it is produced, including standards of software engineering. The use of drones in regulated airspace falls under the authority of the Federal Aviation Administration (FAA).[126] For consumer-facing AI systems, regulation by the Federal Trade Commission (FTC) comes into play. Financial markets using AI technologies, such as in high-frequency trading, come under regulation by the Security Exchange Commission (SEC).

In addition to sector-specific approaches, the somewhat ambiguous and broad regulatory category of "critical infrastructure" may apply to AI applications.[127] The Obama Administration's Presidential Policy Directive (PPD) 21 broadly defines critical infrastructure as composed of "the assets, systems, and networks, whether physical or virtual, so vital to the United States that their incapacitation or destruction would have a debilitating effect on security, national economic security, national public health or safety, or any combination thereof." Today, an enterprise does not come under federal regulation solely by falling under that broad definition. Instead, the general trend of federal policy is to seek regulation in sixteen sectors of the economy.[128]

As regards AI, critical infrastructure is notably defined by the end-user application, and not the technology or sector that actually produces AI software.[129] Software companies such as Google, Facebook, and Amazon have actively lobbied to avoid being designated as critical to the economy, arguing that this would open the door to regulation that would inevitably compromise their rapid product development cycles and ability to innovate.[130] Nonetheless, as the companies creating, operating, and maintaining critical infrastructure use AI, interest will grow in regulating that software.

---

126  FAA controls the ways drones fly, requires drones to be semiautonomous as opposed to autonomous, requires visual connection to the drone, and enforces no-fly zones close to airports.

127  "Presidential Policy Directive (PPD-21)—Critical Infrastructure Security and Resilience," The White House, February 12, 2013, accessed August 1, 2016, https://www.whitehouse.gov/the-press-office/2013/02/12/presidential-policy-directive-critical-infrastructure-security-and-resil.

128  PPD 21 identifies agencies responsible in each case. Chemical: Department of Homeland Security; Commercial Facilities: Department of Homeland Security; Communications: Department of Homeland Security; Critical Manufacturing: Department of Homeland Security; Dams: Department of Homeland Security; Defense Industrial Base: Department of Defense; Emergency Services: Department of Homeland Security; Energy: Department of Energy; Financial Services: Department of the Treasury; Food and Agriculture: U.S. Department of Agriculture and Department of Health and Human Services; Government Facilities: Department of Homeland Security and General Services Administration; Healthcare and Public Health: Department of Health and Human Services; Information Technology: Department of Homeland Security; Nuclear Reactors, Materials, and Waste: Department of Homeland Security; Transportation Systems: Department of Homeland Security and Department of Transportation; Water and Wastewater Systems: Environmental Protection Agency.

129  In "ICYMI- Business Groups Urge White House to Rethink Cyber Security Order," *Internet Association*, March 5, 2013, accessed August 1, 2016, https://internetassociation.org/030513gov-3/: "Obama's Feb. 12 order says the government can't designate 'commercial information technology products' or consumer information technology services as critical U.S. infrastructure targeted for voluntary computer security standards,' … 'Obama's order isn't meant to get down to the level of products and services and dictate how those products and services behave,' said David LeDuc, senior director of public policy for the Software & Information Industry Association, a Washington trade group that lobbied for the exclusions."

Some existing regulatory regimes for software safety (for example, the FDA's regulation of high consequence medical software) require specific software engineering practices at the developer level. However, modern software systems are often assembled from library components which may be supplied by multiple vendors, and are relatively application-independent. It doesn't seem feasible or desirable to subject all such developers to the standards required for the most critical, rare applications. Nor does it seem advisable to allow unregulated use of such components in safety critical applications. Tradeoffs between promoting innovation and regulating for safety are difficult ones, both conceptually and in practice. At a minimum, regulatory entities will require greater expertise going forward in order to understand the implications of standards and measures put in place by researchers, government, and industry.[131]

## Policy and legal considerations

While a comprehensive examination of the ways artificial intelligence (AI) interacts with the law is beyond the scope of this inaugural report, this much seems clear: as a transformative technology, AI has the potential to challenge any number of legal assumptions in the short, medium, and long term. Precisely how law and policy will adapt to advances in AI—and how AI will adapt to values reflected in law and policy—depends on a variety of social, cultural, economic, and other factors, and is likely to vary by jurisdiction.

American law represents a mixture of common law, federal, state, and local statutes and ordinances, and—perhaps of greatest relevance to AI—regulations. Depending on its instantiation, AI could implicate each of these sources of law. For example, Nevada passed a law broadly permitting autonomous vehicles and instructed the Nevada Department of Motor Vehicles to craft requirements. Meanwhile, the National Highway Transportation Safety Administration has determined that a self-driving car system, rather than the vehicle occupants, can be considered the "driver" of a vehicle. Some car designs sidestep this issue by staying in autonomous mode only when hands are on the wheel (at least every so often), so that the human driver has ultimate control and responsibility. Still, Tesla's adoption of this strategy did not prevent the first traffic fatality involving an autonomous car, which occurred in June of 2016. Such incidents are sure to influence public attitudes towards autonomous driving. And as most people's first experience with embodied agents, autonomous transportation will strongly influence the public's perception of AI.

Driverless cars are, of course, but one example of the many instantiations of AI in services, products, and other contexts. The legal effect of introducing AI into the provision of tax advice, automated trading on the stock market, or generating medical diagnoses will also vary in accordance to the regulators that govern these contexts and the rules that apply within them. Many other examples of AI applications fall within current non-technology-specific policy, including predictive

> **Absent sufficient technical expertise to assess safety or other metrics, national or local officials may refuse to permit a potentially promising application—or green light a sensitive application that has not been adequately vetted.**

---

130  Eric Engleman, "Google Exception in Obama's Cyber Order Questioned as Unwise Gap," *Bloomberg Technology*, March 4, 2013, accessed August 1, 2016, http://www.bloomberg.com/news/articles/2013-03-05/google-exception-in-obama-s-cyber-order-questioned-as-unwise-gap.
131  Ryan Calo, "The Case for a Federal Robotics Commission," *Brookings Report*, September 15, 2014, accessed August 1, 2016, http://www.brookings.edu/research/reports2/2014/09/case-for-federal-robotics-commission.

> As AI applications engage in behavior that, were it done by a human, would constitute a crime, courts and other legal actors will have to puzzle through whom to hold accountable and on what theory.

policing, non-discriminatory loans, healthcare applications such as eldercare and drug delivery, systems designed to interact with children (for example, autonomous tutoring systems are required to respect laws in regard to balanced handling of evolution vs. intelligent design), and interactive entertainment.

Given the present structure of American administrative law, it seems unlikely that AI will be treated comprehensively in the near term. Nevertheless, it is possible to enumerate broad categories of legal and policy issues that AI tends to raise in various contexts.

**Privacy**

Private information about an individual can be revealed through decisions and predictions made by AI. While some of the ways that AI implicates privacy mirror those of technologies such as computers and the internet, other issues may be unique to AI. For example, the potential of AI to predict future behavior based on previous patterns raises challenging questions. Companies already use machine learning to predict credit risk. And states run prisoner details through complex algorithms to predict the likelihood of recidivism when considering parole. In these cases, it is a technical challenge to ensure that factors such as race and sexual orientation are not being used to inform AI-based decisions. Even when such features are not directly provided to the algorithms, they may still correlate strongly with seemingly innocuous features such as zip code. Nonetheless, with careful design, testing, and deployment, AI algorithms may be able to make less biased decisions than a typical person.

Anthropomorphic interfaces increasingly associated with AI raise novel privacy concerns. Social science research suggests people are hardwired to respond to anthropomorphic technology as though it were human. Subjects in one study were more likely to answer when they were born if the computer first stated when it was built.[132] In another, they skipped sensitive questions when posed by an anthropomorphic interface.[133] At a basic level lies the question: Will humans continue to enjoy the prospect of solitude in a world permeated by apparently social agents "living" in our houses, cars, offices, hospital rooms, and phones?[134]

**Innovation policy**

Early law and policy decisions concerning liability and speech helped ensure the commercial viability of the Internet. By contrast, the software industry arguably suffers today from the decision of firms to pivot from open and free software to the more aggressive pursuit of intellectual property protections, resulting in what some have termed patent "thickets." Striking the proper balance between incentivizing innovation in AI while promoting cooperation and protection against third party harm will prove a central challenge.

**Liability (civil)**

As AI is organized to directly affect the world, even physically, liability for harms caused by AI will increase in salience. The prospect that AI will behave in ways designers do not expect challenges the prevailing assumption within tort law that courts only compensate for foreseeable injuries. Courts might arbitrarily assign liability to a human actor even when liability is better located elsewhere for reasons of fairness or efficiency. Alternatively, courts could refuse to find liability because the defendant before the court did not, and could not, foresee the harm that the AI caused. Liability would then fall by default on the blameless victim. The role of product liability—and the responsibility that falls to companies manufacturing these products— will likely grow when human actors become less responsible for the actions of a machine.

---

132   Youngme Moon, "Intimate Exchanges: Using Computers to Elicit Self-Disclosure from Consumers," *Journal of Consumer Research* 26, no. 4 (March 2000): 323–339.

133   Lee Sproull, Mani Subramani, Sara Kiesler , Janet H. Walker, and Keith Waters, "When the Interface is a Face," *Human-Computer Interaction* 11, no. 2 (1996): 97–124.

134   M. Ryan Calo, "People Can Be So Fake: A New Dimension to Privacy and Technology Scholarship," *Penn State Law Review* 114, no. 3 (2010): 809–855.

**Liability (criminal)**

If tort law expects harms to be foreseeable, criminal law goes further to expect that harms be *intended*. US law in particular attaches great importance to the concept of *mens rea*—the intending mind. As AI applications engage in behavior that, were it done by a human, would constitute a crime, courts and other legal actors will have to puzzle through whom to hold accountable and on what theory.

**Agency**

The above issues raise the question of whether and under what circumstances an AI system could operate as the agent of a person or corporation. Already regulatory bodies in the United States, Canada, and elsewhere are setting the conditions under which software can enter into a binding contract.[135] The more AI conducts legally salient activities, the greater the challenge to principles of agency under the law.

**Certification**

The very notion of "artificial intelligence" suggests a substitution for human skill and ingenuity. And in many contexts, ranging from driving to performing surgery or practicing law, a human must attain some certification or license before performing a given task. Accordingly, law and policy will have to—and already does—grapple with how to determine competency in an AI system. For example, imagine a robotics company creates a surgical platform capable of autonomously removing an appendix. Or imagine a law firm writes an application capable of rendering legal advice. Today, it is unclear from a legal perspective *who* in this picture would have to pass the medical boards or legal bar, let alone *where* they would be required to do so.[136]

**Labor**

As AI substitutes for human roles, some jobs will be eliminated and new jobs will be created. The net effect on jobs is ambiguous, but labor markets are unlikely to benefit everyone evenly. The demand for some types of skills or abilities will likely drop significantly, negatively affecting the employment levels and wages of people with those skills.[137] While the ultimate effects on income levels and distribution are not inevitable, they depend substantially on government policies, on the way companies choose to organize work, and on decisions by individuals to invest in learning new skills and seeking new types of work and income opportunities. People who find their employment altered or terminated as a consequence of advances of AI may seek recourse in the legislature and courts. This may be why Littler Mendelson LLP— perhaps the largest employment law firm in the world—has an entire practice group to address robotics and artificial intelligence.

**Taxation**

Federal, state, and local revenue sources may be affected. Accomplishing a task using AI instead of a person can be faster and more accurate—and avoid employment taxes. As a result, AI applications could increasingly shift investment from payroll and income to capital expenditure. Depending on a state budget's reliance on payroll and income tax, such a shift could be destabilizing. AI may also display different

> AI applications could increasingly shift investment from payroll and income to capital expenditure. Depending on a state budget's reliance on payroll and income tax, such a shift could be destabilizing.

---

135   Ian R. Kerr, "Ensuring the Success of Contract Formation in Agent-Mediated Electronic Commerce," *Electronic Commerce Research* 1 (2001): 183–202.
136   Ryan Calo, "Digital Agenda's public discussion on 'The effects of robotics on economics, labour and society,'" *Ausschuss Digitale Agenda*, (Deutsche Bundestag: Ausschussdrucksache A-Drs. 18(24)102), June 22, 2016, accessed August 1, 2016, https://www.bundestag.de/blob/428266/195a1cde8d5347849accbbe60ed91865/a-drs-18-24-102-data.pdf.
137   Brynjolfsson and McAfee, "Race Against the Machine: How the Digital Revolution is Accelerating Innovation, Driving Productivity, and Irreversibly Transforming Employment and the Economy (2011)"; Brynjolfsson and McAfee, *Second Machine Age*.

> Like other technologies, AI has the potential to be used for good or nefarious purposes. A vigorous and informed debate about how to best steer AI in ways that enrich our lives and our society is an urgent and vital need.

"habits" than people, resulting in still fewer revenue sources. The many municipalities relying on income from speeding or parking tickets will have to find alternatives if autonomous cars can drop people off and find distance parking, or if they are programmed not to violate the law. As a result, government bodies trying to balance their budgets in light of advances in AI may pass legislation to slow down or alter the course of the technology.

**Politics**

AI technologies are already being used by political actors in gerrymandering and targeted "robocalls" designed to suppress votes, and on social media platforms in the form of "bots."[138] They can enable coordinated protest as well as the ability to predict protests, and promote greater transparency in politics by more accurately pinpointing who said what, when. Thus, administrative and regulatory laws regarding AI can be designed to promote greater democratic participation or, if ill-conceived, to reduce it.

This list is not exhaustive and focuses largely on domestic policy in the United States, leaving out many areas of law that AI is likely to touch. One lesson that might be drawn concerns the growing disconnect between the context-specific way in which AI is governed today and a wider consideration of themes shared by AI technologies across industries or sectors of society. It could be tempting to create new institutional configurations capable of amassing expertise and setting AI standards across multiple contexts. The Study Panel's consensus is that attempts to regulate "AI" in general would be misguided, since there is no clear definition of AI (it isn't any one thing), and the risks and considerations are very different in different domains. Instead, policymakers should recognize that to varying degrees and over time, various industries will need distinct, appropriate, regulations that touch on software built using AI or incorporating AI in some way. The government will need the expertise to scrutinize standards and technology developed by the private and public sector, and to craft regulations where necessary.

### Guidelines for the future

Faced with the profound changes that AI technologies can produce, pressure for "more" and "tougher" regulation is inevitable. Misunderstanding about what AI is and is not, especially against a background of scare-mongering, could fuel opposition to technologies that could benefit everyone. This would be a tragic mistake. Regulation that stifles innovation, or relocates it to other jurisdictions, would be similarly counterproductive.

Fortunately, principles that guide successful regulation of current digital technologies can be instructive. A recent multi-year study comparing privacy regulation in four European countries and the United States, for example, yielded counter-intuitive results.[139] Those countries, such as Spain and France, with strict and detailed regulations bred a "compliance mentality" within corporations, which had the effect of discouraging both innovation and robust privacy protections. Rather than taking responsibility for privacy protection internally and developing a professional staff to foster it in business and manufacturing processes, or engaging with privacy advocates or academics outside their walls, these companies viewed privacy as a compliance activity. Their focus was on avoiding fines or punishments, rather than proactively designing technology and adapting practices to protect privacy.

By contrast, the regulatory environment in the United States and Germany, which combined more ambiguous goals with tough transparency requirements and meaningful enforcement, were more successful in catalyzing companies to view privacy as their responsibility. Broad legal mandates encouraged companies to develop a professional staff and processes to enforce privacy controls, engage with outside stakeholders, and to adapt their practices to technology advances. Requiring greater transparency enabled civil society groups and media to become credible enforcers both in court and in the court of public opinion, making privacy more salient to corporate boards and leading them to further invest in privacy protection.

---

138  Political Bots, accessed August 1, 2016, http://politicalbots.org/.
139  Kenneth A. Bamberger and Deirdre K. Mulligan, *Privacy on the Ground: Driving Corporate Behavior in the United States and Europe* (Cambridge, Massachusetts: MIT Press, 2015).

In AI, too, regulators can strengthen a virtuous cycle of activity involving internal and external accountability, transparency, and professionalization, rather than narrow compliance. As AI is integrated into cities, it will continue to challenge existing protections for values such as privacy and accountability. Like other technologies, AI has the potential to be used for good or nefarious purposes. This report has tried to highlight the potential for both. A vigorous and informed debate about how to best steer AI in ways that enrich our lives and our society, while encouraging creativity in the field, is an urgent and vital need. Policies should be evaluated as to whether they democratically foster the development and equitable sharing of AI's benefits, or concentrate power and benefits in the hands of a fortunate few. And since future AI technologies and their effects cannot be foreseen with perfect clarity, policies will need to be continually re-evaluated in the context of observed societal challenges and evidence from fielded systems.

As this report documents, significant AI-related advances have already had an impact on North American cities over the past fifteen years, and even more substantial developments will occur over the next fifteen. Recent advances are largely due to the growth and analysis of large data sets enabled by the Internet, advances in sensory technologies and, more recently, applications of "deep learning." In the coming years, as the public encounters new AI applications in domains such as transportation and healthcare, they must be introduced in ways that build trust and understanding, and respect human and civil rights. While encouraging innovation, policies and processes should address ethical, privacy, and security implications, and should work to ensure that the benefits of AI technologies will be spread broadly and fairly. Doing so will be critical if Artificial Intelligence research and its applications are to exert a positive influence on North American urban life in 2030 and beyond.

## APPENDIX I: A SHORT HISTORY OF AI

This Appendix is based primarily on Nilsson's book[140] and written from the prevalent current perspective, which focuses on data intensive methods and big data. However important, this focus has not yet shown itself to be the solution to all problems. A complete and fully balanced history of the field is beyond the scope of this document.

The field of Artificial Intelligence (AI) was officially born and christened at a 1956 workshop organized by John McCarthy at the Dartmouth Summer Research Project on Artificial Intelligence. The goal was to investigate ways in which machines could be made to simulate aspects of intelligence—the essential idea that has continued to drive the field forward. McCarthy is credited with the first use of the term "artificial intelligence" in the proposal he co-authored for the workshop with Marvin Minsky, Nathaniel Rochester, and Claude Shannon.[141] Many of the people who attended soon led significant projects under the banner of AI, including Arthur Samuel, Oliver Selfridge, Ray Solomonoff, Allen Newell, and Herbert Simon.

Although the Dartmouth workshop created a unified identity for the field and a dedicated research community, many of the technical ideas that have come to characterize AI existed much earlier. In the eighteenth century, Thomas Bayes provided a framework for reasoning about the **probability** of events.[142] In the nineteenth century, George Boole showed that **logical reasoning**—dating back to Aristotle—could be performed *systematically* in the same manner as solving a system of equations.[143] By the turn of the twentieth century, progress in the experimental sciences had led to the emergence of the field of **statistics**,[144] which enables inferences to be drawn rigorously from data. The idea of physically engineering a machine to execute sequences of instructions, which had captured the imagination of pioneers such as Charles Babbage, had matured by the 1950s, and resulted in the construction of the first **electronic computers**.[145] Primitive **robots**, which could sense and act autonomously, had also been built by that time.[146]

The most influential ideas underpinning computer science came from Alan Turing, who proposed a formal model of computing. Turing's classic essay, *Computing Machinery and Intelligence*,[147] imagines the possibility of computers created for simulating intelligence and explores many of the ingredients now associated with AI, including how intelligence might be tested, and how machines might automatically *learn*. Though these ideas inspired AI, Turing did not have access to the computing resources needed to translate his ideas into action.

Several focal areas in the quest for AI emerged between the 1950s and the 1970s.[148]

> The field of Artificial Intelligence (AI) was officially born and christened at a 1956 workshop. The goal was to investigate ways in which machines could be made to simulate aspects of intelligence—the essential idea that has continued to drive the field forward.

---

140  Nilsson, *The Quest for Artificial Intelligence*.
141  J. McCarthy, Marvin L. Minsky, Nathaniel Rochester, and Claude E. Shannon, "A Proposal for the Dartmouth Summer Research Project on Artificial Intelligence," August 31, 1955, accessed August 1, 2016, http://www-formal.stanford.edu/jmc/history/dartmouth/dartmouth.html.
142  Thomas Bayes, "An Essay towards Solving a Problem in the Doctrine of Chances," *Philosophical Transactions of the Royal Society of London* 53 (January 1, 1763): 370–418, accessed August 1, 2016, http://rstl.royalsocietypublishing.org/search?fulltext=an+essay+towards+solving&submit=yes&andorexactfulltext=and&x=0&y=0.
143  George Boole, *An Investigation of the Laws of Thought on Which are Founded the Mathematical Theories of Logic and Probabilities*, (Macmillan, 1854, reprinted with corrections, Dover Publications, New York, NY, 1958, and reissued by Cambridge University Press, 2009), accessed August 1, 2016, http://ebooks.cambridge.org/ebook.jsf?bid=CBO9780511693090.
144  "History of statistics," *Wikipedia*, Last modified June 3, 2016, accessed August 1, 2016, https://en.wikipedia.org/wiki/History_of_statistics.
145  Joel N. Shurkin, *Engines of the Mind: The Evolution of the Computer from Mainframes to Microprocessors* (New York: W. W. Norton & Company, 1996).
146  William Grey Walter, "An Electromechanical Animal," *Dialectica* 4 (1950): 42–49.
147  A. M. Turing, "Computing Machinery and Intelligence," *Mind* 59, no. 236 (1950): 433–460.
148  Marvin Minsky, "Steps toward Artificial Intelligence," MIT Media Laboratory, October 24, 1960, accessed August 1, 2016, http://web.media.mit.edu/~minsky/papers/steps.html.

Newell and Simon pioneered the foray into **heuristic search**, an efficient procedure for finding solutions in large, combinatorial spaces. In particular, they applied this idea to construct proofs of mathematical theorems, first through their Logic Theorist program, and then through the General Problem Solver.[149] In the area of **computer vision**, early work in character recognition by Selfridge and colleagues[150] laid the basis for more complex applications such as face recognition.[151] By the late sixties, work had also begun on **natural language processing**.[152] "Shakey", a wheeled robot built at SRI International, launched the field of **mobile robotics**. Samuel's Checkers-playing program, which improved itself through self-play, was one of the first working instances of a **machine learning** system.[153] Rosenblatt's *Perceptron*,[154] a computational model based on biological neurons, became the basis for the field of **artificial neural networks**. Feigenbaum and others advocated [155]the case for building **expert systems**—knowledge repositories tailored for specialized domains such as chemistry and medical diagnosis.[156]

Early conceptual progress assumed the existence of a symbolic system that could be reasoned about and built upon. But by the 1980s, despite this promising headway made into different aspects of artificial intelligence, the field still could boast no significant *practical* successes. This gap between theory and practice arose in part from an insufficient emphasis within the AI community on *grounding* systems physically, with direct access to environmental signals and data. There was also an overemphasis on Boolean (True/False) logic, overlooking the need to quantify uncertainty. The field was forced to take cognizance of these shortcomings in the mid-1980s, since interest in AI began to drop, and funding dried up. Nilsson calls this period the "AI winter."

A much needed resurgence in the nineties built upon the idea that "Good Old-Fashioned AI"[157] was inadequate as an end-to-end approach to building intelligent systems. Rather, intelligent systems needed to be built from the ground up, at all times *solving* the task at hand, albeit with different degrees of proficiency.[158] Technological progress had also made the task of building systems driven by real-world data more feasible. Cheaper and more reliable hardware for sensing and actuation made robots easier to build. Further, the Internet's capacity for gathering large amounts of data, and the availability of computing power and storage to process that data, enabled statistical techniques that, by design, derive solutions from data. These developments have allowed AI to emerge in the past two decades as a profound influence on our daily lives, as detailed in Section II.

> **Although the separation of AI into sub-fields has enabled deep technical progress along several different fronts, synthesizing intelligence at any reasonable scale invariably requires many different ideas to be integrated.**

In summary, following is a list of some of the traditional sub-areas of AI. As described in Section II, some of them are currently "hotter" than others for various reasons. But that is neither to minimize the historical importance of the others, nor to say that they may not re-emerge as hot areas in the future.

- Search and Planning deal with reasoning about goal-directed behavior. Search plays a key role, for example, in chess-playing programs such as Deep Blue, in deciding which move (behavior) will ultimately lead to a win (goal).
- The area of Knowledge Representation and Reasoning involves processing information (typically when in large amounts) into a structured form that can be queried more reliably and efficiently. IBM's Watson program, which beat human contenders to win the Jeopardy challenge in 2011, was largely based on an efficient scheme for organizing, indexing, and retrieving large amounts of information gathered from various sources.[159]
- Machine Learning is a paradigm that enables systems to automatically improve their performance at a task by observing relevant data. Indeed, machine learning has been the key contributor to the AI surge in the past few decades, ranging from search and product recommendation engines, to systems for speech recognition, fraud detection, image understanding, and countless other tasks that once relied on human skill and judgment. The automation of these tasks has enabled the scaling up of services such as e-commerce.
- As more and more intelligent systems get built, a natural question to consider is how such systems will interact with each other. The field of Multi-Agent Systems considers this question, which is becoming increasingly important in on-line marketplaces and transportation systems.
- From its early days, AI has taken up the design and construction of systems that are embodied in the real world. The area of Robotics investigates fundamental aspects of sensing and acting—and especially their integration—that enable a robot to behave effectively. Since robots and other computer systems share the living world with human beings, the specialized subject of Human Robot Interaction has also become prominent in recent decades.
- Machine perception has always played a central role in AI, partly in developing robotics, but also as a completely independent area of study. The most commonly studied perception modalities are Computer Vision and Natural Language Processing, each of which is attended to by large and vibrant communities.
- Several other focus areas within AI today are consequences of the growth of the Internet. Social Network Analysis investigates the effect of neighborhood relations in influencing the behavior of individuals and communities. Crowdsourcing is yet another innovative problem-solving technique, which relies on harnessing human intelligence (typically from thousands of humans) to solve hard computational problems.

Although the separation of AI into sub-fields has enabled deep technical progress along several different fronts, synthesizing intelligence at any reasonable scale invariably requires many different ideas to be integrated. For example, the AlphaGo program[160] [161] that recently defeated the current human champion at the game of Go used multiple machine learning algorithms for training itself, and also used a sophisticated search procedure while playing the game.

---

159  David A. Ferrucci, "Introduction to 'This is Watson,'" *IBM Journal of Research and Development*, 56, no. 3-4 (2012): 1.
160  David Silver et al., "Mastering the Game of Go with Deep Neural Networks and Tree Search."
161  Steven Borowiec and Tracey Lien, "AlphaGo beats human Go champ in milestone for artificial intelligence," *Los Angeles Times*, March 12, 2016, accessed August 1, 2016, http://www.latimes.com/world/asia/la-fg-korea-alphago-20160312-story.html.